\documentclass[10pt,letterpaper]{article}
\pdfoutput=1
\usepackage{xcolor}
\usepackage{array}
\usepackage{geometry}
\usepackage{tabularx}

\usepackage{changepage}

\usepackage[utf8]{inputenc}

\usepackage{textcomp,marvosym}

\usepackage{fixltx2e}

\usepackage{amsmath,amssymb}

\usepackage{mathtools}

\usepackage[superscript]{cite}

\usepackage{nameref}

\usepackage{hyperref}
\hypersetup{colorlinks,linkcolor={blue},citecolor={blue},urlcolor={blue}} 

\usepackage[makeroom]{cancel}

\usepackage{breqn}

\usepackage{microtype}
\DisableLigatures[f]{encoding = *, family = * }

\usepackage{rotating}

\usepackage[font=small,aboveskip=1pt,labelfont=bf,labelsep=period,justification=raggedright,singlelinecheck=off]{caption}

\usepackage{float}

\date{}

\usepackage{multirow}

\usepackage{mathtools}
\newcommand{\step}{E_i^{-S_{ij}}}
\newcommand{\pderiv}[2]{\frac{\partial#1}{\partial#2}}
\newcommand{\pdd}[3]{\frac{\partial^2#1}{\partial#2\partial#3}}

\renewcommand{\d}[1]{\ensuremath{\operatorname{d}\!{#1}}}
\newcommand{\x}{\mathbf{x}}
\newcommand{\xss}{\mathbf{x}_{\text{ss}}}
\newcommand{\deriv}[2]{\frac{\text{d}#1}{\text{d}#2}}
\DeclareMathOperator{\Diag}{Diag}
\DeclareMathOperator{\Tr}{Tr}
\newcommand{\defeq}{\vcentcolon=}
\DeclareMathOperator{\Exp}{\text{Exponential}}
\DeclareMathOperator{\Bin}{\text{Binomial}}

\newcommand{\ud}[1]{\textcolor{black}{#1}}

\usepackage{natbib}

\usepackage{framed}

\begin{document}
\vspace*{0.35in}

\begin{flushleft}
{\Large
\textbf\newline{Mitochondrial network state scales mtDNA genetic dynamics}
}
\newline
\\

Juvid Aryaman,$^{\ast, \dagger, \ddagger}$
Charlotte Bowles,$^\S$
Nick S. Jones,$^{\ast, \ast\ast, 1}$
Iain G. Johnston,$^{\dagger\dagger, \S\S, 2}$

\bigskip
$^\ast$ Department of Mathematics, Imperial College London, London, SW7 2AZ, United Kingdom\\
$^\dagger$ Department of Clinical Neurosciences, University of Cambridge, Cambridge, CB2 0QQ, United Kingdom\\
$^\ddagger$ MRC Mitochondrial Biology Unit, University of Cambridge, Cambridge, CB2 0XY, United Kingdom\\
$^\S$ School of Biosciences, University of Birmingham, Birmingham, B15 2TT, United Kingdom\\
$^{\ast\ast}$ EPSRC Centre for the Mathematics of Precision Healthcare, Imperial College London, London, SW7 2AZ, United Kingdom\\
$^{\dagger\dagger}$ Faculty of Mathematics and Natural Sciences, University of Bergen, 5007 Bergen, Norway\\
$^{\S\S}$ Alan Turing Institute, London, NW1 2DB, United Kingdom
\bigskip

$^1$ nick.jones@imperial.ac.uk

$^2$ iain.johnston@uib.no

\bigskip

\end{flushleft}

\raggedbottom

\section*{Abstract}

Mitochondrial DNA (mtDNA) mutations cause severe congenital diseases but may also be associated with healthy aging. MtDNA is stochastically replicated and degraded, and exists within organelles which undergo dynamic fusion and fission. The role of the resulting mitochondrial networks in the time evolution of the cellular proportion of mutated mtDNA molecules (heteroplasmy), and cell-to-cell variability in heteroplasmy (heteroplasmy variance), remains incompletely understood. Heteroplasmy variance is particularly important since it modulates the number of pathological cells in a tissue. Here, we provide the first wide-reaching theoretical framework which bridges mitochondrial network and genetic states. We show that, under a range of conditions, the (genetic) rate of increase in heteroplasmy variance and \textit{de novo} mutation are proportionally modulated by the (physical) fraction of unfused mitochondria, independently of the absolute fission-fusion rate. In the context of selective fusion, we show that intermediate fusion/fission ratios are optimal for the clearance of mtDNA mutants. Our findings imply that modulating network state, mitophagy rate and copy number to slow down heteroplasmy dynamics when mean heteroplasmy is low could have therapeutic advantages for mitochondrial disease and healthy aging. 

\section*{Introduction}

Mitochondrial DNA (mtDNA) encodes elements of the respiratory system vital for cellular function. Mutation of mtDNA is \ud{one of several leading hypotheses} for the cause of normal aging \citep{Lopez13,Kauppila17}, as well as underlying a number of heritable mtDNA-related diseases \citep{Schon12}. Cells typically contain hundreds, or thousands, of copies of mtDNA per cell: each molecule encodes crucial components of the electron transport chain, which generates energy for the cell in the form of ATP. Consequently, the mitochondrial phenotype of a single cell is determined, in part, by its  fluctuating population of mtDNA molecules \citep{Wallace13,Stewart15,Johnston18,Aryaman19}. The broad biomedical implications of mitochondrial DNA mutation, combined with the countable nature of mtDNAs and the stochastic nature of their dynamics, offer the opportunity for mathematical understanding to provide important insights into human health and disease \citep{Aryaman19}.

\begin{figure*}
\begin{center}
\includegraphics[width=\columnwidth]{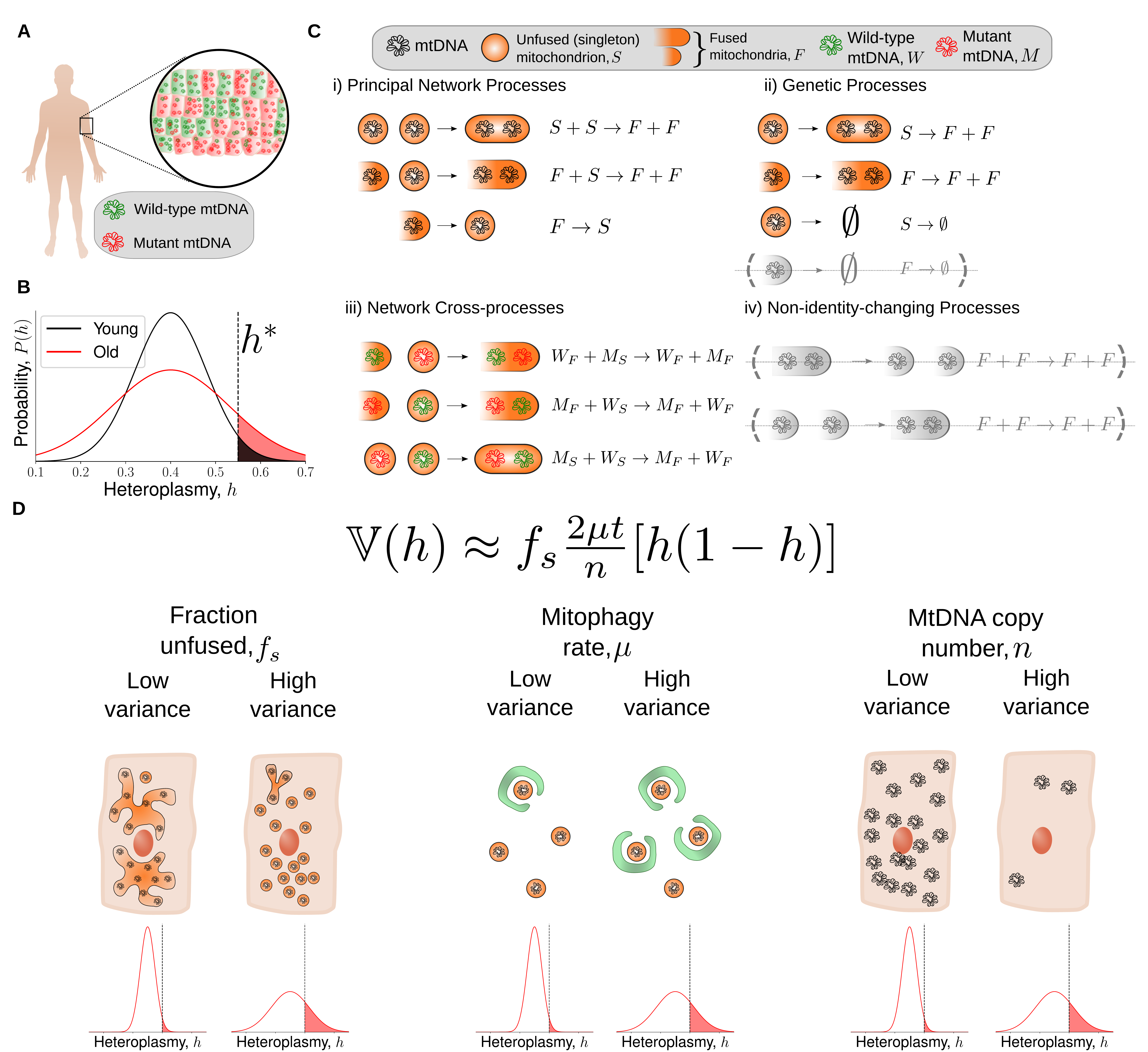}  
\end{center}
\caption{{\bf A simple model bridging mitochondrial networks and genetics yields a wide-reaching, analytically obtained, description of heteroplasmy variance dynamics}. (\textbf{A})~A population of cells from a tissue exhibit inter-cellular heterogeneity in mitochondrial content: both mutant load (heteroplasmy) and copy number. (\textbf{B})~Inter-cellular heterogeneity implies that heteroplasmy is described by a probability distribution. Cells above a threshold heteroplasmy ($h^*$, black dashed line) are thought to exhibit a pathological phenotype. The low-variance distribution (black line) has fewer cells above a pathological threshold heteroplasmy than the high-variance distribution (red line). \ud{Heteroplasmy is depicted as an approximately normal distribution, as this is the regime in which our approximations below hold: i.e.\ when the probability of fixation is small.} (\textbf{C})~ The chemical reaction network we use to model the dynamics of mitochondrial DNA (see Main Text for a detailed description). MtDNAs are assigned a genetic state: mutant ($M$) or wild-type ($W$), and a network state: singleton (i.e.\ unfused, $S$) or fused ($F$). (\textbf{D})~The central result of our work is, assuming that a cell at time $t=0$ is at its (deterministic) steady-state, heteroplasmy variance ($\mathbb{V}(h)$) approximately increases with time ($t$), mitophagy rate ($\mu$) and the fraction of mitochondria that are unfused ($f_s$), and decreases with mtDNA copy number ($n$). Importantly, $\mathbb{V}(h)$ does not depend on the absolute magnitude of the fission-fusion rates. Also see Table~\ref{tab:key_findings} for a summary of our key findings.
}
\label{Fig:net_cartoon}
\end{figure*}

An important observation in mitochondrial physiology is the threshold effect, whereby cells may often tolerate relatively high levels of mtDNA mutation, until the fraction of mutated mtDNAs (termed heteroplasmy) exceeds a certain critical value where a pathological phenotype occurs \citep{Rossignol03,Picard14,Stewart15,Aryaman17b}. Fluctuations within individual cells mean that the fraction of mutant mtDNAs per cell is not constant within a tissue (Figure~\ref{Fig:net_cartoon}A), but follows a probability distribution which changes with time (Figure~\ref{Fig:net_cartoon}B). Here, motivated by a general picture of aging, we will largely focus on the setting of non-dividing cells, which possess two mtDNA variants (although we will also consider \textit{de novo} mutation using simple statistical genetics models). The variance of the distribution of heteroplasmies gives the fraction of cells above a given pathological threshold (Figure~\ref{Fig:net_cartoon}B). Therefore heteroplasmy variance is related to the number of dysfunctional cells \ud{above a phenotypic threshold} within a tissue, and both heteroplasmy mean and variance are directly related to tissue physiology. \ud{Increases in heteroplasmy variance also increase the number of cells below a given threshold heteroplasmy, which can be advantageous in e.g.\ selecting low-heteroplasmy embryos in pre-implantation genetic diagnosis for treating mitochondrial disease \citep{Burgstaller14b, Johnston15}.}

Mitochondria exist within a network which dynamically fuses and fragments. Although the function of mitochondrial networks remains an open question \citep{Hoitzing15}, it is often thought that a combination of network dynamics and mitochondrial autophagy (termed mitophagy) act in concert to perform quality control on the mitochondrial population \citep{Twig08,Johnston18,Aryaman19}. Observations of pervasive intra-mitochondrial mtDNA mutation \citep{Morris17} and universal heteroplasmy in humans \citep{Payne12} suggest that the power of this quality control may be limited. \ud{It has also been suggested that certain mtDNA mutations, such as deletions \citep{Kowald13,Kowald14b,Kowald18} and some point mutations \citep{Samuels13, Ye14, Li15, Lieber19}, are under the influence of selective effects. However, genetic models without selection have proven valuable in explaining the heteroplasmy dynamics both of functional mutations \citep{Elson01,Taylor03, Wonnapinij08} and polymorphisms without dramatic functional consequences \citep{Birky83, Ye14}, and in common cases where mean heteroplasmy shifts are small compared to changes in variances (for instance, in germline development \citep{Johnston15} and post-mitotic tissues \citep{Burgstaller14}). Mean changes seem more likely in high-turnover tissues and when mtDNA variants are genetically distant \citep{Burgstaller14, Pan19}, suggesting that neutral genetic theory may be useful in understanding the dynamics of the set of functionally mild mutations which accumulate during ageing. Neutral genetic theory also provides a valuable null model for understanding mitochondrial genetic dynamics \citep{Chinnery99,Poovathingal09,Johnston16}, potentially allowing us to better understand and quantify when selection is present.} There is thus a set of open questions about how the physical dynamics of mitochondria affect the genetic populations of mtDNA within and between cells under neutral dynamics. 

A number of studies have attempted to understand the impact of the mitochondrial network on mitochondrial dysfunction through computer simulation (reviewed in \cite{Kowald14}). These studies have suggested: that clearance of damaged mtDNA can be assisted by high and funcitonally-selective mitochondrial fusion, or by intermediate fusion and selective mitophagy \citep{Mouli09}; that physical transport of mitochondria can indirectly modulate mitochondrial health through mitochondrial dynamics \citep{Patel13}; that fission-fusion dynamic rates modulate a trade-off between mutant proliferation and removal \citep{Tam13,Tam15}; and that if fission is damaging, decelerating fission-fusion cycles may improve mitochondrial quality \citep{Figge12}.


Despite providing valuable insights, these previous attempts to link mitochondrial genetics and network dynamics, while important for breaking ground, have centered around complex computer simulations, making it difficult to deduce general laws and principles. Here, we address this lack of a general theoretical framework linking mitochondrial dynamics and genetics. We take a simpler approach in terms of our model structure (Figure~\ref{Fig:net_cartoon}C), allowing us to derive explicit, interpretable, mathematical formulae which provide intuitive understanding, and give a direct account for the phenomena which are observed in our model (Figure~\ref{Fig:net_cartoon}D). Our results hold for a range of variant model structures. Simplified approaches using stochastic modelling have shown success in understanding mitochondrial physiology from a purely genetic perspective \citep{Chinnery99,Capps03,Johnston16}. Furthermore, there currently exists limited evidence for pronounced, universal, selective differences of mitochondrial variants \textit{in vivo} \citep{Stewart14,Hoitzing17b}. Our basic approach therefore also differs from previous modelling attempts, since our model is neutral with respect to genetics (no replicative advantage or selective mitophagy) and the mitochondrial network (no selective fusion). Evidence for negative selection of particular mtDNA mutations has been observed \textit{in vivo} \citep{Ye14,Morris17}; we therefore extend our analysis to explore selectivity in the context of mitochondrial quality control using our simplified framework.

Here, we reveal the first general mathematical principle linking (physical) network state and (genetic) heteroplasmy statistics (Figure~\ref{Fig:net_cartoon}D). Our models potentially allow rich interactions between mitochondrial genetic and network dynamics, yet we find that a simple link emerges. For a broad range of situations, the expansion of mtDNA mutants is strongly modulated by network state, such that the rate of increase of heteroplasmy variance, and the rate of accumulation of \textit{de novo} mutation, is proportional to the fraction of unfused mitochondria. We discover that this result stems from the general notion that fusion shields mtDNAs from turnover, since autophagy of large fragments of the mitochondrial network are unlikely, and consequently rescales time. Importantly, we used our model for network dynamics to show that heteroplasmy variance is independent of the absolute magnitude of the fusion and fission rates due to a separation of timescales between genetic and network processes (in contrast to \cite{Tam15}). Surprisingly, we find the dependence of heteroplasmy statistics upon network state arises when the mitochondrial population size is controlled through replication, and vanishes when it is controlled through mitophagy, shedding new light on the physiological importance of the mode of mtDNA control. We show that when fusion is selective, intermediate fusion/fission ratios are optimal for the clearance of mutated mtDNAs (in contrast to \cite{Mouli09}). When mitophagy is selective, complete fragmentation of the network results in the most effective elimination of mitochondrial mutants (in contrast to \cite{Mouli09}). We also confirm that mitophagy and mitochondrial DNA copy number affect the rate of accumulation of \textit{de novo} mutations \citep{Johnston16}, see Table~\ref{tab:key_findings} for a summary of our key findings. We suggest that pharmacological interventions which promote fusion, slow mitophagy and increase copy number earlier in development may slow the rate of accumulation of pathologically mutated cells, with implications for mitochondrial disease and aging.

\section*{Materials and Methods}
\subsection*{Stochastic modelling of the coupling between genetic and network dynamics of mtDNA populations}

Our modelling approach takes a chemical master equation perspective by combining a general model of neutral genetic drift (for instance, see \cite{Chinnery99,Johnston16}) with a model of mitochondrial network dynamics. \ud{We seek to understand the influence of the mitochondrial network upon mitochondrial genetics. The network state itself is influenced by several factors including metabolic poise and the respiratory state of mitochondria \citep{Szabadkai06, Hoitzing15, Mishra16}, which we do not consider explicitly here.} We consider the existence of two mitochondrial alleles, wild-type ($W$) and mutant ($M$), \ud{existing within a post-mitotic cell without cell division, with mtDNAs undergoing turnover (or ``relaxed replication'' \cite{Stewart15})}. MtDNAs exist within mitochondria, which undergo fusion and fission. We therefore assign mtDNAs a network state: fused ($F$) or unfused (we term ``singleton'', $S$). This representation of the mitochondrial network allows us to include the effects of the mitochondrial network in a simple way, without the need to resort to a spatial model or consider the precise network structure, allowing us to make analytic progress and derive interpretable formulae in a more general range of situations. 

Our model can be decomposed into three notional blocks  (Figure~\ref{Fig:net_cartoon}C). Firstly, the principal network processes denote fusion and fission of mitochondria containing mtDNAs of the same allele
\begin{eqnarray}
X_S+ X_S &\xrightarrow{\gamma}& X_F + X_F \label{eq:non-lin-fus}\\
X_F+ X_S &\xrightarrow{\gamma}& X_F + X_F \label{eq:lin-fus}\\
X_F &\xrightarrow{\beta}& X_S \label{eq:fis}
\end{eqnarray}
where $X$ denotes either a wild-type ($W$) or a mutant ($M$) mtDNA (therefore a set of chemical reactions analogous to Eq.~\eqref{eq:non-lin-fus}-\eqref{eq:fis} exist for both DNA species). $\gamma$ and $\beta$ are the stochastic rate constants for fusion and fission respectively. 

Secondly, mtDNAs are replicated and degraded through a set of reactions termed genetic processes. A central assumption is that all degradation of mtDNAs occur through mitophagy, and that only small pieces of the mitochondrial network are susceptible to mitophagy; for parsimony we take the limit of only the singletons being susceptible to mitophagy
\begin{eqnarray}
X_S &\xrightarrow{\lambda}& X_F + X_F \label{eq:birth}\\
X_F &\xrightarrow{\lambda}& X_F + X_F \\
X_S &\xrightarrow{\mu}& \emptyset \label{eq:death_s}
\end{eqnarray}
where $\lambda$ and $\mu$ are the replication and mitophagy rates respectively, which are shared by both $W$ and $M$ resulting in a so-called `neutral' genetic model. Eq.~\eqref{eq:death_s} denotes removal of the species from the system. The effect of allowing non-zero degradation of fused species is discussed in \nameref{sec:Methods} (see Eq.~\eqref{eq:fused_deg_xi} and Figure~\ref{Fig:stoch_extras}E). Replication of a singleton changes the network state of the mtDNA into a fused species, since replication occurs within the same membrane-bound organelle. An alternative model of singletons which replicate into singletons, thereby associating mitochondrial replication with fission \citep{Lewis16}, leaves our central result (Figure~\ref{Fig:net_cartoon}D) unchanged (see \nameref{sec:Methods}, Eq.~\eqref{eq:s_birth_s}). The system may be considered neutral since both $W$ and $M$ possess the same replication and degradation rates per molecule of mtDNA at any instance in time.

Finally, mtDNAs of different genotypes may interact through fusion via a set of reactions we term network cross-processes:
\begin{eqnarray}
W_F + M_S &\xrightarrow{\gamma}& W_F + M_F \label{eq:fus-cross-1} \\ 
M_F + W_S &\xrightarrow{\gamma}& M_F + W_F \label{eq:fus-cross-2} \\
W_S + M_S &\xrightarrow{\gamma}& W_F + M_F. \label{eq:fus-cross-3}
\end{eqnarray}
Any fusion or fission event which does not involve the generation or removal of a singleton leaves our system unchanged; we term such events as non-identity-changing processes, which can be ignored in our system (see \nameref{sec:Methods}, \nameref{SI:rate_renorm} for a discussion of rate renormalization). We have neglected \textit{de novo} mutation in the model description above (although we will consider \textit{de novo} mutation using a modified infinite sites Moran model below).

We found that treating $\lambda=$ const led to instability in total copy number (see \nameref{sec:Methods}, \nameref{SI:const_rate}), which is not credible. We therefore favoured a state-dependent replication rate such that copy number is controlled to a particular value, as has been done by previous authors \citep{Chinnery99,Capps03,Johnston16}. Allowing lower-case variables to denote the copy number of their respective molecular species, we will focus on a linear replication rate of the form \citep{Hoitzing17,Hoitzing17b}:
\begin{equation}
\lambda = \lambda(w_T,m_T) = \mu + b(\kappa - (w_T + \delta m_T)) \label{eq:lin_feedback_ctrl}
\end{equation}
where $w_T = w_s+w_f$ is the total wild-type copy number, and similarly for $m_T$. The lower-case variables $w_s$, $w_f$, $m_s$, and $m_f$ denote the copy numbers of the corresponding chemical species ($W_S$, $W_F$, $M_S$, and $M_F$). $b$ is a parameter which determines the strength with which total copy number is controlled to a target copy number, and $\kappa$ is a parameter which is indicative of (but not equivalent to) the steady state copy number. $\delta$ indicates the relative contribution of mutant mtDNAs to the control strength and is linked to the ``maintenance of wild-type'' hypothesis \citep{Durham07,Stewart15}. When $0 \leq \delta < 1$, and both mutant and wild-type species are present, mutants have a lower contribution to the birth rate than wild-types. When wild-types are absent, the population size will be larger than when there are no mutants: hence mutants have a higher carrying capacity in this regime. We have modelled the mitophagy rate as constant per mtDNA. We do, however, explore relaxing this constraint below by allowing mitophagy to be a function of state, and also affect mutants differentially under quality control. \ud{$\lambda$ may be re-written as $\lambda = k_1 + k_2 w_T + k_3 m_T$ for constants $k_i$, and so only consists of 3 independent parameters. However we will retain $\lambda$ in the form of Eq.~\eqref{eq:lin_feedback_ctrl} since the parameters $\mu$, $b$, $\kappa$, and $\delta$ have the distinct physiological meanings described above \citep{Hoitzing17, Hoitzing17b}. Furthermore, $\lambda$ may in general also depend on other cellular features such as mitochondrial reactive oxygen species. Here, we seek to explain mitochondrial behaviour under a simple set of governing principles, but our approach can naturally be combined with a description of these additional factors to build a more comprehensive model.} Analogues of this model (without a network) have been applied to mitochondrial systems \citep{Chinnery99,Capps03}. Overall, our simple model consists of 4 species ($W_S,W_F,M_S,M_F$), \ud{6 independent} parameters and 15 reactions, and captures the central property that mitochondria fragment before degradation \citep{Twig08}.

\ud{Throughout this work, we define heteroplasmy as the mutant allele fraction per cell of a mitochondrially-encoded variant \citep{Wonnapinij08, Samuels10, Aryaman19}:
\begin{equation}
h(\x)= (m_s+m_f)/(w_s+w_f+m_s+m_f) \label{eq:het_defn}
\end{equation}
where $\x=(w_s,w_f,m_s,m_f)$ is the state of the system (not to be confused with mitochondrial ``respiratory states''). Hence, a heteroplasmy of $h=1$ denotes a cell with 100\% mutant mtDNA (i.e. a homoplasmic cell in the mutant allele). Arguably, ``mutant allele fraction'' would be a more precise description of Eq.\eqref{eq:het_defn} but we retain the use of heteroplasmy for consistency. To convert to a definition of heteroplasmy which is maximal when the mutant allele fraction is 50\%, one may simply use the conversion $0.5-|h(\x)-0.5|$.}

\subsubsection*{Statistical Analysis} 

In Figures \ref{Fig:stoch_extras}B, \ref{Fig:sweeps}A-I, we compare Eq.~\eqref{eq:ansatz} and Eq.~\eqref{eq:ansatz_indep_net} to stochastic simulations, for various parametrizations and replication/degradation rates. To quantify the accuracy of these equations in predicting $\mathbb{V}(h,t)$, we define the following error metric $\epsilon$ 
\begin{equation}
\epsilon = \left| 1-\frac{\dot{\mathbb{V}}(h,t)_{\text{Th}} }{\mathbb{E}_t(\dot{\mathbb{V}}(h,t)_{\text{Sim}})} \right| \label{eq:error_metric}
\end{equation}
where $\dot{\mathbb{V}}(h,t)$ is the time derivative of heteroplasmy variance with subscripts denoting theory (Th) and simulation (Sim). An expectation over time ($\mathbb{E}_t$) is taken for the stochastic simulations, whereas $\dot{\mathbb{V}}(h,t)$ is a scalar quantity for Eq.~\eqref{eq:ansatz} and Eq.~\eqref{eq:ansatz_indep_net}. 

\subsection*{Data Availability}

Code for simulations and analysis can be accessed at \sloppy{\url{https://GitHub.com/ImperialCollegeLondon/MitoNetworksGenetics}}

\section*{Results}
\subsection*{Mitochondrial network state rescales the linear increase of heteroplasmy variance over time, independently of fission-fusion rate magnitudes}

\begin{figure*}
\begin{center}
\includegraphics[width=\columnwidth]{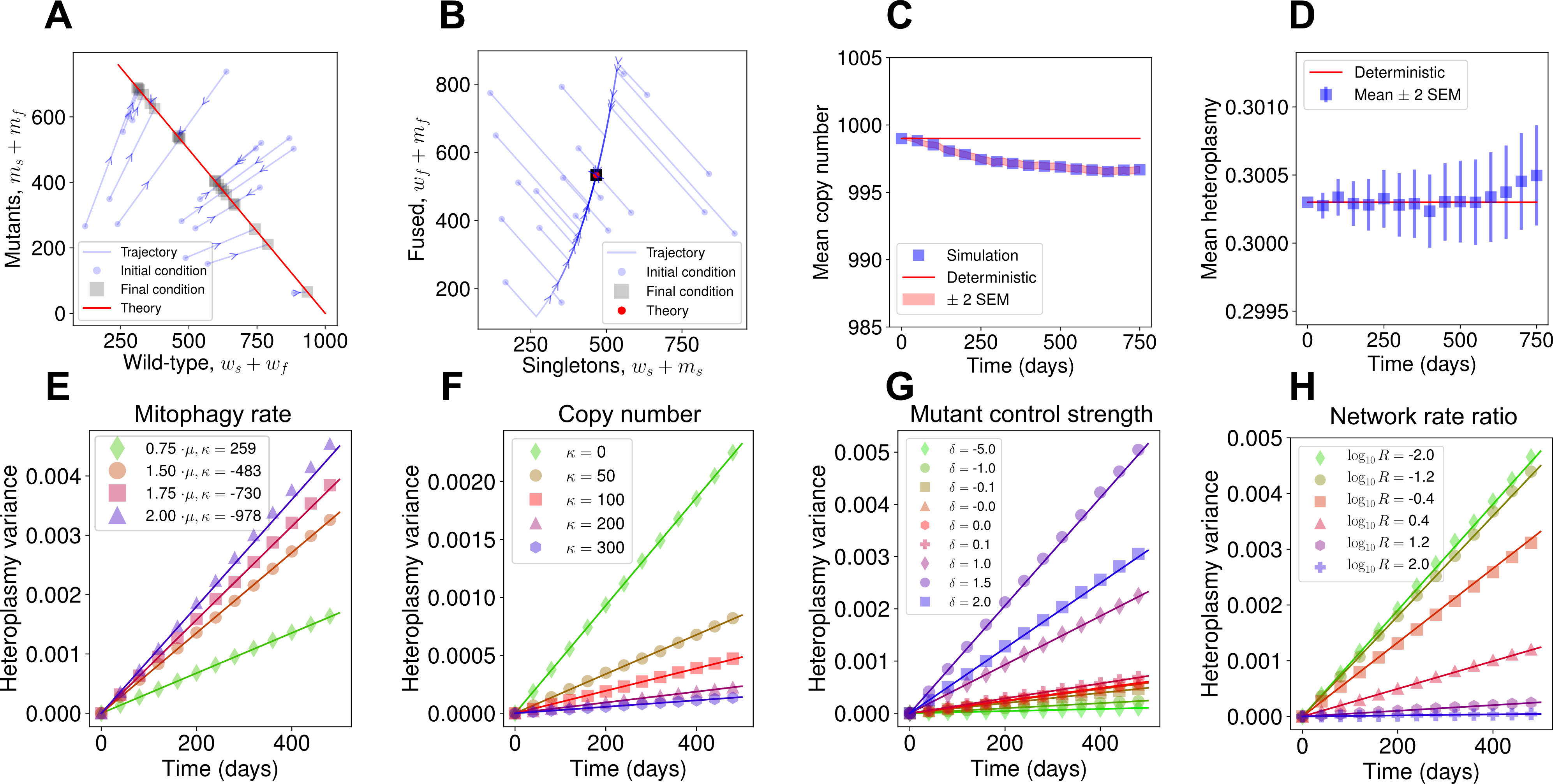}  
\end{center}
\caption{{\bf General mathematical principles linking heteroplasmy variance to network dynamics}. Wild-type and mutant copy numbers (\textbf{A}) and fused and unfused copy numbers (\textbf{B}) both  move towards a line of steady states under a deterministic model, as indicated by arrows. In stochastic simulation, mean copy number (\textbf{C}) is initially slightly perturbed from the deterministic treatment of the system, and then remains constant, while mean heteroplasmy (\textbf{D}) remains invariant with time (see Eq.~\eqref{eq:drift_ss_0}). In (\textbf{E})-(\textbf{H}), we show that Eq.~\eqref{eq:ansatz} holds across many cellular circumstances: lines give analytic results, points are from stochastic simulation. Heteroplasmy variance behaviour is successfully predicted for varying mitophagy rate (\textbf{E}), steady state copy number (\textbf{F}), mutation sensing (\textbf{G}), and fusion rate (\textbf{H}). In (\textbf{H}), fusion and fission rates are redefined as $\gamma \rightarrow \gamma_0 M R$ and $\beta \rightarrow \beta_0 M$ where $M$ and $R$ denote the relative magnitude and ratio of the network rates, and $\gamma_0, \beta_0$ denote the nominal parametrizations of the fusion and fission rates respectively (see Table~\ref{table:params}). Figure~\ref{Fig:stoch_extras}D shows a sweep of $M$ over the same logarithmic range when $R=1$. See Figure \ref{Fig:sweeps}A-I and Table \ref{tab:param_sweeps} for parameter sweeps numerically demonstrating the generality of the result for different mtDNA control modes.
}
\label{Fig:model_expl}
\end{figure*}

We first performed a deterministic analysis of the system presented in Eqs.~\eqref{eq:non-lin-fus}--\eqref{eq:lin_feedback_ctrl}, by converting the reactions into an analogous set of four coupled ordinary differential equations (see Eqs.~\eqref{eq:phi_ws}--\eqref{eq:phi_mf}), and choosing a biologically-motivated approximate parametrization (which we will term the `nominal' parametrization, see \nameref{sec:Methods}, \nameref{SI:nom_param}, and Table~\ref{table:params}). Figures \ref{Fig:model_expl}A-B show that copy numbers of each individual species change in time such that the state approaches a line of steady states (Eqs.~\eqref{eq:det_ss_ws}--\eqref{eq:det_ss_mf}), as seen in other neutral genetic models \citep{Capps03,Hoitzing17b}. Upon reaching this line, total copy number remains constant (Figure \ref{Fig:det_extras}A) and the state of the system ceases to change with time. This is a consequence of performing a deterministic analysis, which neglects stochastic effects, and our choice of replication rate in Eq.~\eqref{eq:lin_feedback_ctrl} which decreases with total copy number when $w_T + \delta m_T > \kappa$ and vice versa, guiding the total population to a fixed total copy number. Varying the fission ($\beta$) and fusion ($\gamma$) rates revealed a negative linear relationship between the steady-state fraction of singletons and copy number (Figure \ref{Fig:det_extras}B).

We may also simulate the system in Eqs.~\eqref{eq:non-lin-fus}--\eqref{eq:fus-cross-3} stochastically, using the stochastic simulation algorithm \citep{Gillespie76}, which showed that mean copy number is slightly perturbed from the deterministic prediction due to the influence of variance upon the mean \citep{Grima11, Hoitzing17b} (Figure \ref{Fig:model_expl}C). The stationarity of total copy number is a consequence of using $\delta=1$ for our nominal parametrization (i.e. the line of steady states is also a line of constant copy number). Choosing $\delta \neq 1$ results in a difference in carrying capacities between the two species, and non-stationarity of mean total copy number, as trajectories spread along the line of steady states to different total copy numbers. Copy number variance initially increases since trajectories are all initialised at the same state, but plateaus because trajectories are constrained in their copy number to remain near the attracting line of steady states (Figure \ref{Fig:stoch_extras}A). Mean heteroplasmy remains constant through time under this model (Figure \ref{Fig:model_expl}D, see \citep{Birky83}). This is unsurprising since each species possesses the same replication and degradation rate, so neither species is preferred.

From stochastic simulations we observed that, for sufficiently short times, heteroplasmy variance increases approximately linearly through time for a range of parametrizations (Figure \ref{Fig:model_expl}E-H), which is in agreement with recent single-cell oocyte measurements in mice \citep{Burgstaller18}. Previous work has also shown a linear increase in heteroplasmy variance through time for purely genetic models of mtDNA dynamics (see \cite{Johnston16}). We sought to understand the influence of mitochondrial network dynamics upon the rate of increase of heteroplasmy variance. 

To this end, we analytically explored the influence of mitochondrial dynamics on mtDNA variability. Assuming  that the state of the system above is initialised at its deterministic steady state ($\x(t=0) = \xss$), we took the limit of limit of large mtDNA copy numbers, fast fission-fusion dynamics, and applied a second-order truncation of the Kramers-Moyal expansion \citep{Gardiner85} to the chemical master equation describing the dynamics of the system (see \nameref{sec:Methods}). This yielded a stochastic differential equation for heteroplasmy, via It\^{o}'s formula \citep{Jacobs10}. Upon forcing the state variables onto the steady-state line \citep{Constable16}, we derived Eq.~\eqref{eq:SDE_network_sys}, which may be approximated for sufficiently short times as 
\begin{equation}
\mathbb{V}(h) \approx \left. f_s(\x) \frac{2 \mu t}{n(\x)} h(\x)(1-h(\x))\right|_{\x=\xss}. \label{eq:ansatz}
\end{equation}
Here, $\mathbb{V}(h)$ is the variance of heteroplasmy, $\mu$ is the mitophagy rate, $n(\x)$ is the total copy number and $f_s(\x)$ is the fraction of unfused (singleton) mtDNAs, and is thus a measure of the fragmentation of the mitochondrial network. $\xss$ is the (deterministic) steady state of the system. Eq.~\eqref{eq:ansatz} demonstrates that \textit{mtDNA heteroplasmy variance increases approximately linearly with time ($t$) at a rate scaled by the fraction of unfused mitochondria, mitophagy rate, and inverse population size}. We find that Eq.~\eqref{eq:ansatz} closely matches heteroplasmy variance dynamics from stochastic simulation, for sufficiently short times after initialisation, for a variety of parametrizations of the system (Figure \ref{Fig:model_expl}E-H, Figure~\ref{Fig:lfc_low_half_life}). 

To our knowledge, Eq.~\eqref{eq:ansatz} reflects the first analytical principle linking mitochondrial dynamics and the cellular population genetics of mtDNA variance. Its simple form allows several intuitive interpretations. As time progresses, replication and degradation of both species occurs, allowing the ratio of species to fluctuate; hence we expect $\mathbb{V}(h)$ to increase with time according to random genetic drift (Figure \ref{Fig:model_expl}E-H). The rate of occurrence of replication/degradation events is set by the mitophagy rate $\mu$, since degradation events are balanced by replication rates to maintain population size; hence, random genetic drift occurs more quickly if there is a larger turnover in the population (Figure \ref{Fig:model_expl}E). We expect $\mathbb{V}(h)$ to increase more slowly in large population sizes, since the birth of e.g. 1 mutant in a large population induces a small change in heteroplasmy (Figure \ref{Fig:model_expl}F). The factor of $h(1-h)$ encodes the state-dependence of heteroplasmy variance, exemplified by the observation that if a cell is initialised at $h=0$ or $h=1$, heteroplasmy must remain at its initial value (since the model above does not consider \textit{de novo} mutation, see below) and so heteroplasmy variance is zero. Furthermore, the rate of increase of heteroplasmy variance is maximal when a cell's initial value of heteroplasmy is 1/2. In Figure~\ref{Fig:model_expl}G, we show that Eq.~\eqref{eq:ansatz} is able to recapitulate the rate of heteroplasmy variance increase across different values of $\delta$, which are hypothesized to correspond to different replicative sensing strengths of different mitochondrial mutations \citep{Hoitzing17b}. We also show in Figures~\ref{Fig:stoch_extras}B\&{}C that Eq.~\eqref{eq:ansatz} is robust to the choice of feedback control strength $b$ in Eq.~\eqref{eq:lin_feedback_ctrl}. \ud{$n(\x)$, $f(\x)$, and $h(\x)$ in Eq.\eqref{eq:ansatz} are not independent degrees of freedom in this model: they are functions of the state vector $\x$, where $\x$ is determined by the parametrization and initial conditions of the model. Hence, the parameter sweeps in Figure \ref{Fig:model_expl}E-H and Figures~\ref{Fig:stoch_extras}B\&{}C also implicitly vary over these functions of state by varying the steady state $\xss$.}

In Eq.~\eqref{eq:death_s}, we have made the important assumption that only unfused mitochondria can be degraded via mitophagy, as seen by \cite{Twig08}, hence the total propensity of mtDNA turnover is limited by the number of mtDNAs which are actually susceptible to mitophagy. Strikingly, we find that \textit{the dynamics of heteroplasmy variance are independent of the absolute rate of fusion and fission, only depending on the fraction of unfused mtDNAs at any particular point in time} (see Figure~\ref{Fig:model_expl}H and Figure~\ref{Fig:stoch_extras}D). This observation, which contrasts with the model of \citep{Tam13,Tam15} (see Discussion), arises from the observation that mitochondrial network dynamics are much faster than replication and degradation of mtDNA, by around a factor of $\beta/\mu \approx 10^3$ (see Table~\ref{table:params}), resulting in the existence of a separation of timescales between network and genetic processes. In the derivation of Eq.~\eqref{eq:ansatz}, we have assumed that fission-fusion rates are infinite, which simplifies $\mathbb{V}(h)$ into a form which is independent of the magnitude of the fission-fusion rate. A parameter sweep of the magnitude and ratio of the fission-fusion rates reveals that, if the fusion and fission rates are sufficiently small, Eq.~\eqref{eq:ansatz} breaks down and $\mathbb{V}(h)$ gains dependence upon the magnitude of these rates (see Figure \ref{Fig:sweeps}A). This regime is, however, for network rates which are approximately 100 times smaller than the biologically-motivated nominal parametrization shown in Figure \ref{Fig:model_expl}A-D where the fission-fusion rate becomes comparable to the mitophagy rate. Since fission-fusion takes place on a faster timescale than mtDNA turnover, we may neglect this region of parameter space as being implausible.

\ud{Eq.~\eqref{eq:ansatz} can be viewed as describing the ``quasi-stationary state'' where the probability of extinction of either allele is negligible \citep{Johnston16}. On longer timescales, or if mtDNA half-life is short \citep{Poovathingal12}, the probability of fixation becomes appreciable. In this case, Eq.~\eqref{eq:ansatz} over-estimates $\mathbf{V}(h)$ as heteroplasmy variance gradually becomes sub-linear with time, see Figure~\ref{Fig:lfc_low_half_life}C\&{}D. This is evident through inspection of Eq.~\eqref{eq:SDE_network_sys}, which shows that cellular trajectories which reach $h=0$ or $h=1$ cease to diffuse in heteroplasmy space, and so heteroplasmy variance cannot increase indefinitely. Consequently, the depiction of heteroplasmy variance in Fig.~\ref{Fig:net_cartoon}B,D as being approximately normally distributed corresponds to the regime in which our approximation holds, and is a valid subset of the behaviours displayed by heteroplasmy dynamics under more sophisticated models (e.g.\ the Kimura distribution \cite{Kimura55, Wonnapinij08}). Further analytical developments may be possible to take into account extinction (e.g.\ see \cite{Wonnapinij08, Assaf10}). However, the linear regime for heteroplasmy variance has been observed to be a substantial component of mtDNA dynamics in e.g.\ mouse oocytes \citep{Burgstaller18}.}

\subsubsection*{The influence of mitochondrial dynamics upon heteroplasmy variance under different models of genetic mtDNA control}

To demonstrate the generality of this result, we explored several alternative forms of cellular mtDNA control \citep{Johnston16}. We found that when copy number is controlled through the replication rate function (i.e.\ $\lambda = \lambda(\x)$, $\mu=$ const), when the fusion and fission rates were high and the fixation probability ($P(h=0)$ or $P(h=1)$) was negligible, Eq.~\eqref{eq:ansatz} accurately described $\mathbb{V}(h)$ across all of the replication rates investigated, see Figure \ref{Fig:sweeps}A-F. The same mathematical argument to show Eq.~\eqref{eq:ansatz} for the replication rate in Eq.~\eqref{eq:lin_feedback_ctrl} may be applied to these alternative replication rates where a closed-form solution for the deterministic steady state may be written down (see \nameref{sec:Methods}, \nameref{SI:ODE_net_deriv}). Interestingly, when copy number is controlled through the degradation rate (i.e.\ $\lambda = $const, $\mu = \mu(\x)$), heteroplasmy variance loses its dependence upon network state entirely and the $f_s$ term is lost from Eq.~\eqref{eq:ansatz} (see Eq.~\eqref{eq:ansatz_indep_net} and Figure \ref{Fig:sweeps}G-I). A similar mathematical argument was applied to reveal how this dependence is lost (see \nameref{sec:Methods}, \nameref{SI:proof_vh_lfc}).

In order to provide an intuitive account for why control in the replication rate, versus control in the degradation rate, determines whether or not heteroplasmy variance has network dependence, we investigated a time-rescaled form of the Moran process (see \nameref{sec:Methods}, \nameref{SI:mod_moran}). The Moran process is structurally much simpler than the model presented above, to the point of being unrealistic, in that the mitochondrial population size is constrained to be constant between consecutive time steps. Despite this, the modified Moran process proved to be insightful. We find that, when copy number is controlled through the replication rate, the absence of death in the fused subpopulation means the timescale of the system (being the time to the next death event) is proportional to $f_s$. In contrast, when copy number is controlled through the degradation rate, the presence of a constant birth rate in the entire population means the timescale of the system (being the time to the next birth event) is independent of $f_s$ (see Eq.~\eqref{eq:rescaled_moran_vh} and surrounding discussion).

\subsection*{Control strategies against mutant expansions} 

In this study, we have argued that the rate of increase of heteroplasmy variance, and therefore the rate of accumulation of pathologically mutated cells within a tissue, increases with mitophagy rate ($\mu$), decreases with total mtDNA copy number per cell ($n$) and increases with the fraction of unfused mitochondria (termed ``singletons'', $f_s$), see Eq.~\eqref{eq:ansatz}. Below, we explore how biological modulation of these variables influences the accumulation of mutations. We use this new insight to propose three classes of strategy to control mutation accumulation and hence address associated issues in aging and disease, and discuss these strategies through the lens of existing biological literature.

\subsubsection*{Targeting network state against mutant expansions}

\begin{figure*}
\begin{center}
\includegraphics[width=0.8\columnwidth]{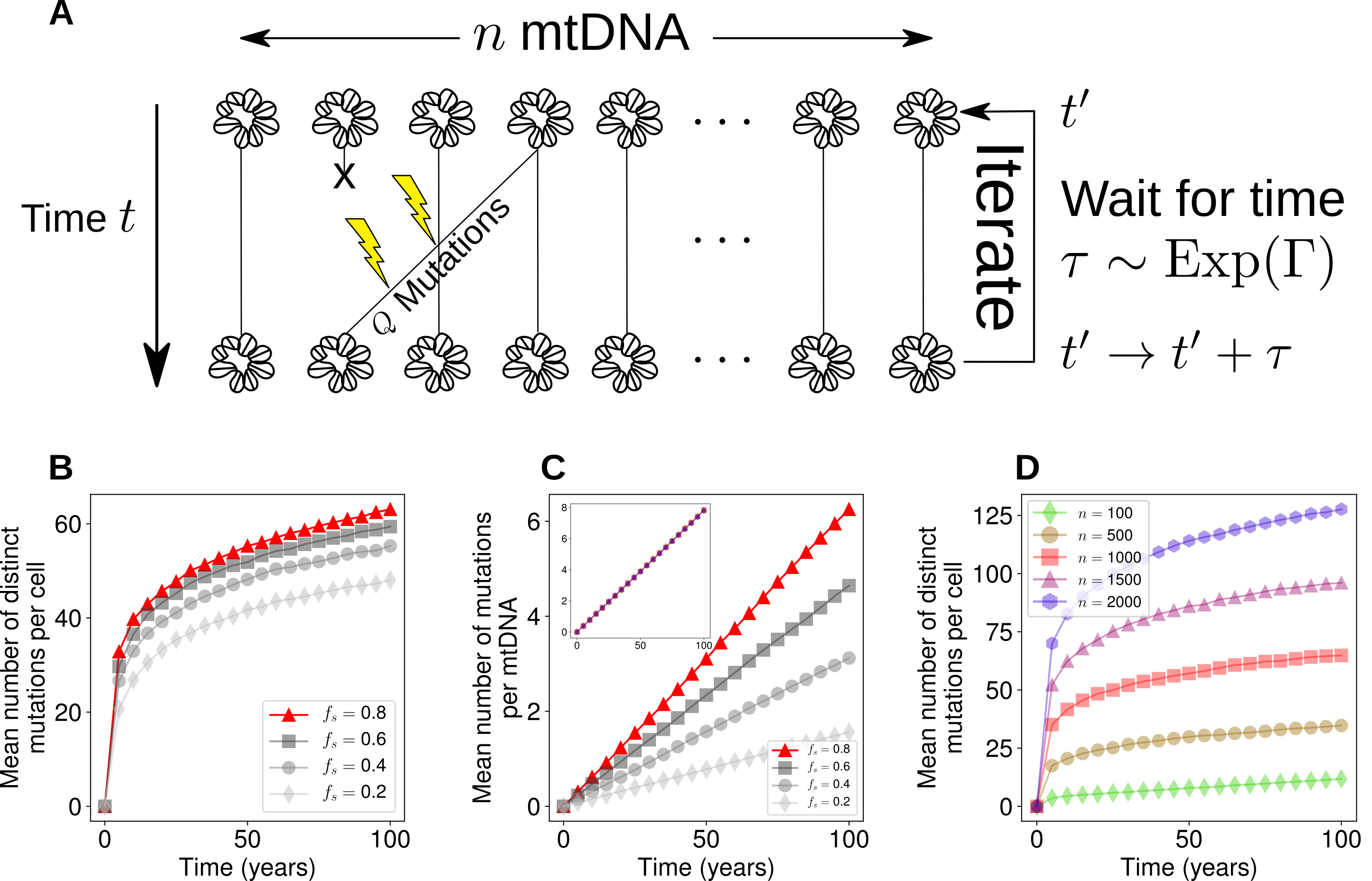}  
\end{center}
\caption{{\bf Rate of \textit{de novo} mutation accumulation is sensitive to the network state/mitophagy rate and copy number for a time-rescaled infinite sites Moran model}. (\textbf{A})~An infinite sites Moran model where $Q$ mutations occur per Moran step (see Eq.~\eqref{eq:inf_sites_Q}). (\textbf{B-D})~Influence of our proposed intervention strategies. (\textbf{B})~Mean number of distinct mutations increases with the fraction of unfused mitochondria. This corresponds to a simple rescaling of time, so all but one of the parametrizations are shown in grey.  (\textbf{C})~ The mean number of mutations per mtDNA also increases with the fraction of unfused mitochondria. Inset shows that the mean number of mutations per mtDNA is independent of the number of mtDNAs per cell; values of $n$ are the same as in (\textbf{D}). (\textbf{D})~Mean number of mutations per cell increases according to the population size of mtDNAs. Standard error in the mean is too small to visualise, so error bars are neglected, given $10^3$ realizations.
}
\label{Fig:inf_sites}
\end{figure*}

In order to explore the role of the mitochondrial network in the accumulation of \textit{de novo} mutations, we invoked an infinite sites Moran model \citep{Kimura69} (see Figure \ref{Fig:inf_sites}A). Single cells were modelled over time as having a fixed mitochondrial copy number ($n$), and at each time step one mtDNA is randomly chosen for duplication and one (which can be the same) for removal. The individual replicated incurs $Q$ \textit{de novo} mutations, where $Q$ is binomially distributed according to
\begin{equation}
Q \sim \Bin(L_{\text{mtDNA}}, \eta) \label{eq:inf_sites_Q}
\end{equation}
where $\Bin(N,p)$ is a binomial random variable with $N$ trials and probability $p$ of success. $L_{\text{mtDNA}} = 16569$ is the length of mtDNA in base pairs and $\eta = 5.6\times 10^{-7}$ is the mutation rate per base pair per doubling \citep{Zheng06}; hence each base pair is idealized to have an equal probability of mutation upon replication. In \nameref{sec:Methods}, Eq.~\eqref{eq:moran_rate_ctrl_rep}, we argue that when population size is controlled in the replication rate, the inter-event rate ($\Gamma$) of the Moran process is effectively rescaled by the fraction of unfused mitochondria, i.e. $\Gamma = \mu n f_s$, which we apply here.

Figure \ref{Fig:inf_sites}B shows that in the infinite sites model, the consequence of Eq.~\eqref{eq:moran_rate_ctrl_rep} is that the rate of accumulation of mutations per cell reduces as the mitochondrial network becomes more fused, as does the mean number of mutations per mtDNA (Figure \ref{Fig:inf_sites}C). These observations are intuitive: since fusion serves to shield the population from mitophagy, mtDNA turnover slows down, and therefore there are fewer opportunities for replication errors to occur per unit time. Different values of $f_s$ in Figures \ref{Fig:inf_sites}B\&{}C therefore correspond to a rescaling of time i.e.\ stretching of the time-axis. \ud{The absolute number of mutations predicted in Figure \ref{Fig:inf_sites}B may over-estimate the true number of mutations per cell (and of course depends on our choice of mutation rate), since a subset of mutations will experience either positive or negative selection. However, quantification of the number of distinct mitochondrial mutants in single cells remains under-explored, as most mutations will have a variant allele fraction close to 0\% or 100\% \citep{Birky83}, which are challenging to measure, especially through bulk sequencing.}

A study by \cite{Chen10} observed the effect of deletion of two proteins which are involved in mitochondrial fusion (Mfn1 and Mfn2) in mouse skeletal muscle. \ud{Although knock-out studies present difficulties in extending their insights into the physiological case, the authors observed that} fragmentation of the mitochondrial network induced severe depletion of mtDNA copy number (\ud{which we also observed in} Figure \ref{Fig:det_extras}B). Furthermore, the authors observed that the number of mutations per base pair increased upon fragmentation, \ud{which we also observed in} the infinite sites model where fragmentation effectively results in a faster turnover of mtDNA (Figure \ref{Fig:inf_sites}C).

Our models predict that promoting mitochondrial fusion has a two-fold effect: firstly, it slows the increase of heteroplasmy variance (see Eq.~\eqref{eq:ansatz} and Figure~\ref{Fig:model_expl}H); secondly, it reduces the rate of accumulation of distinct mutations (see Figure~\ref{Fig:inf_sites}B\&{}C). These two effects are both a consequence of mitochondrial fusion rescaling the time to the next turnover event, and therefore the rate of random genetic drift. As a consequence, this simple model suggests that promoting fusion earlier in development (assuming mean heteroplasmy is low) could slow down the accumulation and spread of mitochondrial mutations, and perhaps slow aging. 


If we assume that fusion is selective in favour of wild-type mtDNAs, which appears to be the case at least for some mutations under therapeutic conditions \citep{Suen10,Kandul16}, we predict that a balance between fusion and fission is the most effective means of removing mutant mtDNAs (see below), perhaps explaining why mitochondrial networks are often observed to exist as balanced between mitochondrial fusion and fission \citep{Sukhorukov12,Zamponi18}. In contrast, if selective mitophagy pathways are induced then promoting fragmentation is predicted to accelerate the clearance of mutants (see below). 

\subsubsection*{Targeting mitophagy rate against mutant expansions}

Alterations in the mitophagy rate $\mu$ have a comparable effect to changes in $f_s$ in terms of reducing the rate of heteroplasmy variance (see Eq.~\eqref{eq:ansatz}) and the rate of \textit{de novo} mutation (Figure~\ref{Fig:inf_sites}B\&{}C) since they both serve to rescale time. Our theory therefore suggests that inhibition of basal mitophagy may be able to slow down the rate of random genetic drift, and perhaps healthy aging, by locking-in low levels of heteroplasmy. Indeed, it has been shown that mouse oocytes \citep{Boudoures17} as well as mouse hematopoietic stem cells \citep{Almeida17} have comparatively low levels of mitophagy, which is consistent with the idea that these pluripotent cells attempt to minimise genetic drift by slowing down mtDNA turnover. A previous modelling study has also shown that mutation frequency increases with mitochondrial turnover \citep{Poovathingal09}. 

Alternatively, it has also been shown that the presence of heteroplasmy, in genotypes which are healthy when present at 100\%, can induce fitness disadvantages \citep{Acton07,Sharpley12,Bagwan18}. In cases where heteroplasmy itself is disadvantageous, especially in later life where such mutations may have already accumulated, accelerating heteroplasmy variance increase to achieve fixation of a species could be advantageous. However, this will not avoid cell-to-cell variability, and the physiological consequences for tissues of such mosaicism is unclear.

\subsubsection*{Targeting copy number against mutant expansions}

To investigate the role of mtDNA copy number (mtCN) on the accumulation of \textit{de novo} mutations, we set $f_s = 1$ such that $\Gamma = \mu n$ (i.e.\ a standard Moran process). We found that varying mtCN did not affect the mean number of mutations per molecule of mtDNA (Figure~\ref{Fig:inf_sites}C, inset). However, as the population size becomes larger, the total number of distinct mutations increases accordingly (Figure~\ref{Fig:inf_sites}D). In contrast to our predictions, a recent study by \cite{Wachsmuth16} found a negative correlation between mtCN and the number of distinct mutations in skeletal muscle. However, \cite{Wachsmuth16} also found a correlation between the number of distinct mutations and age, in agreement with our model. Furthermore, the authors used partial regression to find that age was more explanatory than mtCN in explaining the number of distinct mutations, suggesting age as a confounding variable to the influence of copy number. Our work shows that, in addition to age and mtCN, turnover rate and network state also influence the proliferation of mtDNA mutations. Therefore, one would ideally account for these four variables for jointly, in order to fully constrain our model.

A study of single neurons in the substantia nigra of healthy human individuals found that mtCN increased with age \citep{Dolle16}. Furthermore, mice engineered to accumulate mtDNA deletions through faulty mtDNA replication \citep{Trifunovic04} display compensatory increases in mtCN \citep{Perier13}, which potentially explains the ability of these animals to resist neurodegeneration. It is possible that the observed increase in mtCN in these two studies is an adaptive response to slow down random genetic drift (see Eq.~\eqref{eq:ansatz}). In contrast, mtCN reduces with age in skeletal muscle \citep{Wachsmuth16}, as well as in a number of other tissues such as pancreatic islets \citep{Cree08} and peripheral blood cells \citep{Mengel14}. Given the beneficial effects of increased mtCN in neurons, long-term increases in mtCN could delay other age-related pathological phenotypes.

\subsection*{Optimal mitochondrial network configurations for mitochondrial quality control}

Whilst the above models of mtDNA dynamics are neutral (i.e.\ $m$ and $w$ share the same replication and degradation rates), it is often proposed that damaged mitochondria may experience a higher rate of degradation \citep{Narendra08,Kim07}. There are two principal ways in which selection may occur on mutant species. Firstly, mutant mitochondria may be excluded preferentially from the mitochondrial network in a background of unbiased mitophagy. If this is the case, mutants would be unprotected from mitophagy for longer periods of time than wild-types, and therefore be at greater hazard of degradation. We can alter the fusion rate ($\gamma$) in the mutant analogues of Eq.~\eqref{eq:non-lin-fus},\eqref{eq:lin-fus} and Eqs.~\eqref{eq:fus-cross-1}--\eqref{eq:fus-cross-3} by writing
\begin{equation}
\gamma \rightarrow \gamma/(1+\epsilon_f) \label{eq:sel_fus_qc}
\end{equation}
for all fusion reactions involving 1 or more mutant mitochondria where $\epsilon_f > 0$. The second potential selective mechanism we consider is selective mitophagy. In this case, the degradation rate of mutant mitochondria is larger than wild-types, i.e.\ we modify the mutant degradation reaction to
\begin{equation}
M_s \xrightarrow{\mu(1+\epsilon_m)} \emptyset \label{eq:sel_deg_qc}
\end{equation}
for $\epsilon_m > 0$.

\begin{figure*}
\begin{center}
\includegraphics[width=0.8\columnwidth]{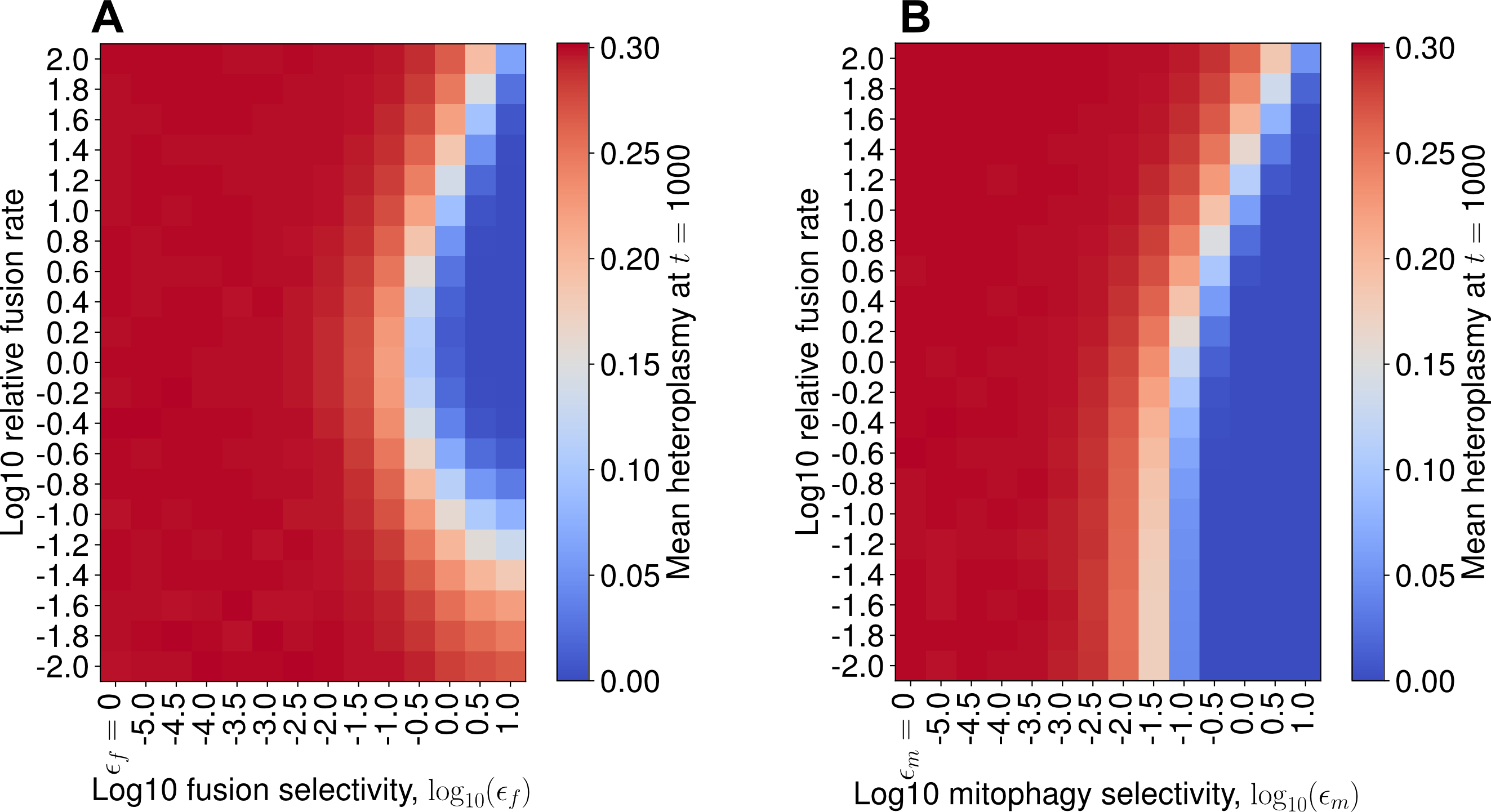}  
\end{center}
\caption{{\bf Selective fusion implies intermediate fusion rates are optimal for mutant clearance whereas selective mitophagy implies complete fission is optimal}. Numerical exploration of the shift in mean heteroplasmy for varying fusion/fission ratio, across different selectivity strengths. Stochastic simulations for mean heteroplasmy, evaluated at 1000 days, with an initial condition of $h=0.3$ and $n=1000$; the state was initialised on the steady state line for the case of $\epsilon_f=\epsilon_m=0$, for $10^4$ iterations. (\textbf{A})~For selective fusion (see Eq.~\eqref{eq:sel_fus_qc}), for each value of fusion selectivity ($\epsilon_f$), the fusion rate ($\gamma$) was varied relative to the nominal parametrization (see Table \ref{table:params}). When $\epsilon_f>0$, the largest reduction in mean heteroplasmy occurs at intermediate values of the fusion rate; a deterministic treatment reveals this to be true for all fusion selectivities investigated (see Figure~\ref{Fig:QC_det_sweep}). (\textbf{B})~For selective mitophagy (see Eq.~\eqref{eq:sel_deg_qc}), when mitophagy selectivity $\epsilon_m >0$, a lower mean heteroplasmy is achieved, the lower the fusion rate (until mean heteroplasmy $=$ 0 is achieved). Hence, complete fission is the optimal strategy for selective mitophagy.
}
\label{Fig:qc}
\end{figure*}

In these two settings, we explore how varying the fusion rate for a given selectivity ($\epsilon_f$ and $\epsilon_m$) affects the extent of reduction in mean heteroplasmy. Figure \ref{Fig:qc}A shows that, \textit{in the context of selective fusion ($\epsilon_f>0$) and non-selective mitophagy ($\epsilon_m=0$) the optimal strategy for clearance of mutants is to have an intermediate fusion/fission ratio}. This was observed for all fusion selectivities investigated (see Figure~\ref{Fig:QC_det_sweep}) Intuitively, if the mitochondrial network is completely fused then, due to mitophagy only acting upon smaller mitochondrial units, mitophagy cannot occur -- so mtDNA turnover ceases and heteroplasmy remains at its initial value. In contrast, if the mitochondrial network completely fissions, there is no mitochondrial network to allow the existence of a quality control mechanism: both mutants and wild-types possess the same probability per unit time of degradation, so mean heteroplasmy does not change. Since both extremes result in no clearance of mutants, the optimal strategy must be to have an intermediate fusion/fission ratio.

In contrast, in Figure \ref{Fig:qc}B, \textit{in the context of non-selective fusion ($\epsilon_f=0$) and selective mitophagy ($\epsilon_m>0$), the optimal strategy for clearance of mutants is to completely fission the mitochondrial network}. Intuitively, if mitophagy is selective, then the more mtDNAs which exist in fragmented organelles, the greater the number of mtDNAs which are susceptible to selective mitophagy, the greater the total rate of selective mitophagy, the faster the clearance of mutants. 

\section*{Discussion} \label{sec:discussion}

In this work, we have sought to unify our understanding of three aspects of mitochondrial physiology -- the mitochondrial network state, mitophagy, and copy number -- with genetic dynamics. The principal virtue of our modelling approach is its simplified nature, which makes general, analytic, quantitative insights available for the first time. In using parsimonious models, we are able to make the first analytic link between the mitochondrial network state and heteroplasmy dynamics. This is in contrast to other computational studies in the field, whose structural complexity make analytic progress difficult, and accounting for their predicted phenomena correspondingly more challenging.

Our bottom-up modeling approach allows for potentially complex interactions between the physical (network) and genetic mitochondrial states of the cell, yet a simple connection emerged from our analysis. We found, for a wide class of models \ud{of post-mitotic cells}, that the rate of linear increase of heteroplasmy variance is  modulated in proportion to the fraction of unfused mitochondria (see Eq.~\eqref{eq:ansatz}). The general notion that mitochondrial fusion shields mtDNAs from turnover, and consequently serves to rescale time, emerges from our analysis. This rescaling of time only holds when mitochondrial copy numbers are controlled through a state-dependent replication rate, and vanishes if copy numbers are controlled through a state-dependent mitophagy rate. We have presented the case of copy number control in the replication rate as being a more intuitive model than control in the degradation rate. The former has the interpretation of biogenesis being varied to maintain a constant population size, with all mtDNAs possessing a characteristic lifetime. The latter has the interpretation of all mtDNA molecules being replicated with a constant probability per unit time, regardless of how large or small the population size is, and changes in mitophagy acting to regulate population size. Such a control strategy seems wasteful in the case of stochastic fluctuations resulting in a population size which is too large, and potentially slow if fluctuations result in a population size which is too small. Furthermore, control in the replication rate means that the mitochondrial network state may act as an additional axis for the cell to control heteroplasmy variance (Figure~\ref{Fig:model_expl}) and the rate of accumulation of \textit{de novo} mutations (Figure \ref{Fig:inf_sites}B\&{}C). Single-mtDNA tracking through confocal microscopy in conjunction with mild mtDNA depletion could shed light on whether the probability of degradation per unit time per mtDNA varies when mtDNA copy number is perturbed, and therefore provide evidence for or against these two possible control strategies.

Our observations provide a substantial change in our understanding of mitochondrial genetics, as it suggests that the mitochondrial network state, in addition to mitochondrial turnover and copy number, must be accounted for in order to predict the rate of spread of mitochondrial mutations in a cellular population. Crucially, through building a model that incorporates mitochondrial dynamics, we find that the dynamics of heteroplasmy variance is independent of the absolute rate of fission-fusion events, since network dynamics occur approximately $10^3$ times faster than mitochondrial turnover, inducing a separation of timescales. The independence of the absolute rate of network dynamics makes way for the possibility of gaining information about heteroplasmy dynamics via the mitochondrial network, \textit{without the need to quantify absolute fission-fusion rates} (for instance through confocal micrographs to quantify the fraction of unfused mitochondria). By linking with classical statistical genetics, we find that the mitochondrial network also modulates the rate of accumulation of \textit{de novo} mutations, also due to the fraction of unfused mitochondria serving to rescale time. We find that, in the context of mitochondrial quality control through selective fusion, an intermediate fusion/fission ratio is optimal due to the finite selectivity of fusion. This latter observation perhaps provides an indication for the reason why we observe mitochondrial networks in an intermediate fusion state under physiological conditions \citep{Sukhorukov12,Zamponi18}.

We have, broadly speaking, considered neutral models of mtDNA genetic dynamics. It is, however, typically suggested that increasing the rate of mitophagy promotes mtDNA quality control, and therefore shrinks the distribution of heteroplasmies towards 0\% mutant (see Eq.~\eqref{eq:sel_fus_qc} and Eq.~\eqref{eq:sel_deg_qc}). If mitophagy is able to change mean heteroplasmy, then a neutral genetic model appears to be inappropriate, as mutants experience a higher rate of degradation. Stimulation of the PINK1/Parkin pathway has been shown to select against deleterious mtDNA mutations \textit{in vitro} \citep{Suen10} and \textit{in vivo} \citep{Kandul16}, as has repression of the mTOR pathway via treatment with rapamycin \citep{Dai13,Kandul16}. However, the necessity of performing a genetic/pharmacological intervention to clear mutations via this pathway suggests that the ability of tissues to selectively remove mitochondrial mutants under physiological conditions is weak. Consequently, neutral models such as our own are useful in understanding how the distribution of heteroplasmy evolves through time under physiological conditions. Indeed, it has been recently shown that mitophagy is basal \citep{Mcwilliams16} and can proceed independently of PINK1 \textit{in vivo} \citep{Mcwilliams18}, perhaps suggesting that mitophagy has non-selective aspects -- although this is yet to be verified conclusively.

\ud{We have paid particular attention to the case of post-mitotic tissues, since these tissues are important for understanding the role of mitochondrial mutations in healthy aging \citep{Khrapko09,Kauppila17}. A typical rate of increase of heteroplasmy variance predicted by Eq.\eqref{eq:ansatz} given our nominal parametrization (Table~\ref{table:params}) is $V'(h)/t=\mathbb{V}(h)/(\mathbb{E}(h)(1-\mathbb{E}(h))t) = 2\mu f_s/n \approx 2.3\times 10^{-5}$~day$^{-1}$ ($f_s=0.5$, $n=1000$). This value accounts for the accumulation of heteroplasmy variance which is attributable to turnover of the mitochondrial population in a post-mitotic cell. However, in the most general case, cell division is also able to induce substantial heteroplasmy variance. For example, $V'(h)/t$ has been measured in model organism germlines to be approximately $9\times 10^{-4}$~day$^{-1}$ in \textit{Drosophila} \citep{Solignac87, Johnston16}, $9\times 10^{-4}$~day$^{-1}$ in NZB/BALB mice \citep{Wai08, Wonnapinij08, Johnston16}, and $2\times10^{-4}$~day$^{-1}$ in single LE and HB mouse oocytes \citep{Burgstaller18}. We see that these rates of increase in heteroplasmy variance are approximately an order of magnitude larger than predictions from our model of purely quiescent turnover, given our nominal parametrisation. Whilst larger mitophagy rates may also potentially induce larger values for $V'(h)/t$ (see \cite{Poovathingal12}, and Figure~\ref{Fig:lfc_low_half_life}C, corrsponding to $V'(h)/t \approx 3.5\times10^{-4}$~ day$^{-1}$) it is clear that partitioning noise (or ``vegetative segregation'', \cite{Stewart15}) is also an important source of variance in heteroplasmy dynamics \citep{Johnston15}. Quantification of heteroplasmy variance in quiescent tissues remains an under-explored area, despite its importance in understanding healthy ageing \citep{Kauppila17, Aryaman19}.}

Our findings reveal some apparent differences with previous studies which link mitochondrial genetics with network dynamics (see Table~\ref{tab:comparison_models}). Firstly, \cite{Tam13,Tam15} found that slower fission-fusion dynamics resulted in larger increases in heteroplasmy variance with time, in contrast to Eq.~\eqref{eq:ansatz} which only depends on fragmentation state and not absolute network rates. The simulation approach of \cite{Tam13,Tam15} allowed for mitophagy to act on whole mitochondria, where mitochondria consist of multiple mtDNAs. Faster fission-fusion dynamics tended to form heteroplasmic mitochondria whereas slower dynamics formed homoplasmic mitochondria. It is intuitive that mitophagy of a homoplasmic mitochondrion induces a larger shift in heteroplasmy than mitophagy of a single mtDNA, hence slower network dynamics form more homoplasmic mitochondria. However, this apparent difference with our findings can naturally be resolved if we consider the regions in parameter space where the fission-fusion rate is much larger than the mitophagy rate, as is empirically observed to be the case \citep{Cagalinec13,Burgstaller14}. If the fission-fusion rates are sufficiently large to ensure heteroplasmic mitochondria, then further increasing the fission-fusion rate is unlikely to have an impact on heteroplasmy dynamics. Hence, this finding is potentially compatible with our study, although future experimental studies investigating intra-mitochondrial heteroplasmy would help constrain these models. \cite{Tam15} also found that fast fission-fusion rates could induce an increase in mean heteroplasmy, in contrast to Figure~\ref{Fig:model_expl}D which shows that mean heteroplasmy is constant with time after a small initial transient due to stochastic effects. We may speculate that the key difference between our treatment and that of \cite{Tam13,Tam15} is the inclusion of cellular subcompartments which induces spatial effects which we do not consider here. The uncertainty in accounting for the phenomena observed in such complex models highlights the virtues of a simplified approach which may yield interpretable laws and principles through analytic treatment.  

The study of \cite{Mouli09} suggested that, in the context of selective fusion, higher fusion rates are optimal. This initially seems to contrast with our finding which states that intermediate fusion rates are optimal for the clearance of mutants (Figure~\ref{Fig:qc}A). However, the high fusion rates in that study do not correspond directly to the highly fused state in our study. Fission automatically follows fusion in \citep{Mouli09}, ensuring at least partial fragmentation, and the high fusion rates for which they identify optimal clearing are an order of magnitude lower than the highest fusion rate they consider. In the case of complete fusion, mitophagy cannot occur in the model of \cite{Mouli09}, so there is no mechanism to remove dysfunctional mitochondria. It is perhaps more accurate to interpret the observations of \cite{Mouli09} as implying that selective fusion shifts the optimal fusion rate higher, when compared to the case of selective mitophagy alone. Therefore the study of \cite{Mouli09} is compatible with Figure~\ref{Fig:qc}A. Furthermore, \cite{Mouli09} also found that when fusion is non-selective and mitophagy is selective, intermediate fusion rates are optimal whereas Figure~\ref{Fig:qc}B shows that complete fragmentation is optimal for clearance of mutants. Optimality of intermediate fusion in the context of selective mitophagy in the model of \cite{Mouli09} likely stems from two aspects of their model: i) mitochondria consist of several units which may or may not be functional; ii) the sigmoidal relationship between number of functional units per mitochondrion and mitochondrial `activity' (the metric by which optimality is measured). Points (i) and (ii) imply that small numbers of dysfunctional mitochondrial units have very little impact on mitochondrial activity, so fusion may boost total mitochondrial activity in the context of small amounts of mutation. So whilst Figure~\ref{Fig:qc}B remains plausible in light of the study of \cite{Mouli09} if reduction of mean heteroplasmy is the objective of the cell, it is also plausible that non-linearities in mitochondrial output under cellular fusion \citep{Hoitzing15} result in intermediate fusion being optimal in terms of energy output in the context of non-selective fusion and selective mitophagy. Future experimental studies quantifying the importance of selective mitophagy under physiological conditions would be beneficial for understanding heteroplasmy variance dynamics. The ubiquity of heteroplasmy \citep{ Payne12,Ye14,Morris17} suggests that a neutral drift approach to mitochondrial genetics may be justified, which contrasts with the studies of \cite{Tam13,Tam15} and \cite{Mouli09} which focus purely on the selective effects of mitochondrial networks. 

In order to fully test our model, further single-cell longitudinal studies are required. For instance, the study by \cite{Burgstaller18} found a linear increase in heteroplasmy variance through time in single oocytes. Our work here has shown that measurement of the network state, as well as turnover and copy number, are required to account for the rate of increase in heteroplasmy variance. Joint longitudinal measurements of $f_s$, $\mu$ and $n$, with heteroplasmy quantification, would allow verification of Eq.~\eqref{eq:ansatz} and aid in determining the extent to which neutral genetic models are explanatory. This could be achieved, for instance, using the mito-QC mouse \citep{Mcwilliams16} which allows visualisation of mitophagy and mitochondrial architecture \textit{in vivo}. Measurement of $f_s$, $\mu$ and $n$, followed by e.g. destructive single-cell whole-genome sequencing of mtDNA would allow validation of how $\mu$, $n$ and $f_s$ influence $\mathbb{V}(h)$ and the rate of \textit{de novo} mutation (see Figure \ref{Fig:inf_sites}). One difficulty is sequencing errors induced through e.g.\ PCR, which hampers our ability accurately measure mtDNA mutation within highly heterogeneous samples \citep{Woods18}. \cite{Morris17} have suggested that single cells are highly heterogeneous in mtDNA mutation, with each mitochondrion possessing 3.9 single-nucleotide variants on average. Error correction strategies during sequencing may pave the way towards high-accuracy mtDNA sequencing \citep{Woods18,Salk18}, and allow us to better constrain models of heteroplasmy dynamics.

\subsection*{Acknowledgments}

We would like to thank Hanne Hoitzing, Thomas McGrath, Abhishek Deshpande, and Ferdinando Insalata for useful discussions. Simulations were performed using the Imperial College High Performance Computing Service. J.A. acknowledges grant support from the BBSRC (BB/J014575/1), and the MRC
Mitochondrial Biology Unit (MC\textunderscore{}UP\textunderscore{}1501/2). N.S.J. acknowledges grant support from the BHF (RE/13/2/30182) and EPSRC (EP/N014529/1). I.G.J. acknowledges funding from ERC StG 805046 (EvoConBiO) and a Turing Fellowship from the Alan Turing Institute.

\newpage

\section*{Supporting Information}\label{sec:Methods}

\setcounter{equation}{0}
\renewcommand{\theequation}{S\arabic{equation}}
\renewcommand\thefigure{S\arabic{figure}}
\setcounter{figure}{0}
\renewcommand\thetable{S\arabic{table}}
\setcounter{table}{0}

\section*{Constant rates yield unstable copy numbers for a model describing mtDNA genetic and network dynamics}\label{SI:const_rate}

We explored a simpler network system than the one presented in the Main Text, but found that it produced instability in mtDNA copy numbers, which we regard as biologically undesirable. Consider the following set of Poisson processes for singleton ($s$) and fused ($f$) species
\begin{eqnarray}
s + s &\xrightarrow{\gamma}& f + f \label{eq:si_fus1}\\ 
s + f &\xrightarrow{\gamma}& f + f\label{eq:si_fus2}\\
f &\xrightarrow{\beta}& s \label{eq:si_fis1}\\
s &\xrightarrow{\alpha \rho}& s + s \label{eq:si_birth_s} \\
f &\xrightarrow{\alpha \rho}& f + f \label{eq:si_birth_f}\\
s &\xrightarrow{\eta \rho}& \emptyset \label{eq:si_death_s}\\
f &\xrightarrow{\rho}& \emptyset  \label{eq:si_death_f}
\end{eqnarray}
where Eq.~\eqref{eq:si_fus1}-\eqref{eq:si_fis1} are analogous to Eq.~\eqref{eq:non-lin-fus}-\eqref{eq:fis} where mutant species are neglected. \ud{Eq.~\eqref{eq:si_birth_s} and \eqref{eq:si_birth_f} are simple birth processes with a shared constant rate $\alpha \rho$. Eq.~\eqref{eq:si_death_s} and \eqref{eq:si_death_f} are simple death processes with rates $\eta \rho$ and $\rho$ respectively.} The parameter $\rho$ is shared amongst all of the birth and death reactions in Eqs.~\eqref{eq:si_birth_s}--\eqref{eq:si_death_f}. $\rho$ represents the intuitive assumption that, in order for a stable population size to exist, birth should balance death. However, for the network to have any effect at all, singletons should be at an increased risk of mitophagy relative to fused species. We represent the increased risk of singleton mitophagy with the parameter $\eta$. Since additional death is introduced into the system when $\eta > 1$, we include the parameter $\alpha>1$ as an increased global biogenesis rate to balance the increased mitophagy of singletons.

We may write the above system as a set of ordinary differential equations
\begin{eqnarray}
\deriv{s}{t} &=& -\gamma s^2 - \gamma fs + \beta f + \alpha \rho s - \eta \rho s \label{eq:si_single_genet_sdot} \\
\deriv{f}{t} &=& \gamma s^2 + \gamma fs - \beta f + \alpha \rho f - \rho f \label{eq:si_single_genet_fdot}
\end{eqnarray}
where we have enforced the stochastic reaction rate to be equivalent to the deterministic reaction rate, and hence the $s^2$ term is proportional to $\gamma$ rather than $2\gamma$ (justification of this is presented below, see Eq.~\eqref{eq:rc_reln}). In Figure~\ref{Fig:justification} we see that the system displays a trivial steady state at $s=f=0$ and a non-trivial steady state. Computing the eigenvalues of the Jacobian matrix at the non-trivial steady state indicates that it is a saddle node, and therefore unstable. Initial copy numbers which are too small tend towards extinction with time, and initial copy numbers which are too large tend towards a copy number explosion. This simple example suggests that a system of this form with constant reaction rates is unstable, and therefore biologically unlikely to exist under reasonable circumstances. We hence consider analogous biochemical reaction networks with a replication rate which is a function of state, to prevent extinction and divergence of the total population size.

\section*{Conversion of a chemical reaction network into ordinary differential equations}\label{SI:CRN_to_ODE}

The following section outlines the steps in converting a set of chemical reactions into a set of ordinary differential equations (ODEs). In particular, we pay special attention to the fact that the rate of a chemical reaction with a stochastic treatment is not always equivalent to the rate in a deterministic treatment \citep{Wilkinson11}, as we will explain below. This subtlety is sometimes overlooked in the literature. This section draws on a number of standard texts \citep{Gillespie76,Van92,Gillespie07, Wilkinson11} as well as \cite{Grima10}. We hope this harmonized treatment will be of help as a future reference.

Consider a general chemical system consisting of $N$ distinct chemical species ($X_i$) interacting via $R$ chemical reactions, where the $j$\textsuperscript{th} reaction is of the form
\begin{equation}
s_{1j}X_1+\dots+s_{Nj}X_N \xrightarrow{\hat{k}_j} r_{1j} X_1+\dots+r_{Nj}X_N \label{eq:gen_chem_sys}
\end{equation}
where $s_{ij}$ and $r_{ij}$ are stoichiometric coefficients. We define $\hat{k}_j$ as the \textit{microscopic} rate for this reaction. The dimensionality of this parameter will vary depending upon the stoichiometric coefficients $s_{ij}$. $\hat{k}_j$ may be loosely interpreted as setting the characteristic timescale (i.e.\ the cross section \citep{Wilkinson11}) of reaction $j$. 

The chemical master equation (CME) describes the dynamics of the joint distribution of the state of the system and time, moving forwards through time. Defining the state of the system as $\mathbf{x}=(x_1,\dots,x_N)^T$, where $x_i$ is the copy number of the $i$\textsuperscript{th} species, allows us to write the CME as \citep{Grima10}
\begin{equation}
\frac{\partial P(\mathbf{x},t|\mathbf{x}_0,t_0)}{\partial t} = \Omega \sum_{j=1}^R \left( \prod_{i=1}^N \step -1\right)\hat{f}_j(\mathbf{x},\Omega)P(\mathbf{x},t|\mathbf{x}_0,t_0) \label{Eq:CME}
\end{equation}
where $\Omega$ is the volume of the compartment in which the reactions occur (also known as the system size), $S_{ij} = r_{ij}-s_{ij}$ is the stoichiometry matrix, and $\step$ is referred to as the step operator and is defined through the relation $\step(g(\mathbf{x})) = g(x_1,\dots,x_i-S_{ij},\dots,x_N)$, for any function of state $g(\mathbf{x})$. $\hat{f}_j(\mathbf{x}, \Omega)$ is the microscopic rate function of reaction $j$, which in general depends on both the state and the system size. A factor of $\Omega$ is explicitly included in this definition of the chemical master equation so that our treatment is compatible with Van Kampen's system size expansion \citep{Van92}.  As a consequence of this, the probability that, given the current state $\mathbf{x}$, the $j$\textsuperscript{th} reaction occurs in the time interval $[t,t+\text{d}t)$ somewhere in $\Omega$ \citep{Gillespie07} is  
\begin{equation}
\hat{a}_j(\mathbf{x},\Omega)\text{d}t \defeq \Omega\hat{f}_j(\mathbf{x},\Omega)\text{d}t. \label{eq:propensity}
\end{equation}
$\hat{a}_j(\mathbf{x},\Omega)$ is termed the propensity function (or ``hazard'') and is of particular relevance in the stochastic simulation algorithm \citep{Gillespie76}, since $\hat{a}_j(\mathbf{x},\Omega)/\sum_j\hat{a}_j(\mathbf{x},\Omega)$ determines the probability that the $j$\textsuperscript{th} reaction occurs next.

For the microscopic rate function, we may write
\begin{equation}
\hat{f}_j(\mathbf{x},\Omega) = \hat{k}_j \prod_{i=1}^N \Omega^{-s_{ij}} \binom{x_i}{s_{ij}}.  \label{eq:micro_rate_fn}
\end{equation}
This equation counts the number of available combinations of reacting molecules \citep{Gillespie76,Wilkinson11}, whilst taking into account scaling with system size \citep{Grima10}.

We also introduce the deterministic rate equation (generally considered to be the macroscopic analogue of the CME) which is defined as \citep{Van92,Grima10}
\begin{equation}
\deriv{\phi_i}{t} = \sum_{j=1}^R S_{ij} \tilde{f}_j(\boldsymbol{\phi}) \label{Eq:RE}
\end{equation}
where $\boldsymbol{\phi}=(\phi_1, \dots, \phi_N)^T$ is the vector of macroscopic concentrations (of dimensions molecules per unit volume) and $\tilde{f}_j(\boldsymbol{\phi})$ is the macroscopic rate function satisfying
\begin{equation}
\tilde{f}_j(\boldsymbol{\phi}) = \tilde{k}_j \prod_{i=1}^N \phi_i^{s_{ij}} \label{eq:macro_rate_fn}
\end{equation}
where $\tilde{k}_j$ is the \textit{macroscopic} rate for the $j$\textsuperscript{th} reaction. We distinguish between $\hat{k}_j$ and $\tilde{k}_j$, respectively the rate constants for the discrete and continuous pictures, although this distinction is sometimes not emphasized in the literature \citep{Grima10, Grima11, Van92}. The physical meaning of $\tilde{k}_j$ is not immediately obvious: we argue that this parameter only gains physical meaning through the following procedure.

As stated by \cite{Wilkinson11}, if we intend for the microscopic description in Eq.~\eqref{Eq:CME} to correspond to the macroscopic description in Eq.~\eqref{Eq:RE}, the rate of consumption/production of particles for every reaction must be the same in the deterministic limit of the stochastic system (the conditions for which we define below). Therefore, we apply the following constraint in the limit of large copy numbers
\begin{equation}
\lim_{x_i \rightarrow \infty} \hat{f}_j(\mathbf{x}, \Omega) = \tilde{f}_j(\boldsymbol{\phi})  \quad \forall\ i,j. \label{eq:equiv_rates_RE_CME}
\end{equation}
In applying this constraint on all species $i$ and all reactions $j$, we may derive a general relationship between $\hat{k}_j$ and $\tilde{k}_j$
\begin{equation}
\lim_{x_i \rightarrow \infty} \hat{k}_j \prod_{i=1}^N \Omega^{-s_{ij}} \frac{x_i!}{s_{ij}!(x_i-s_{ij})!} = \tilde{k}_j \prod_{i=1}^N \phi_i^{s_{ij}}. \label{eq:exact_rc_reln}
\end{equation}
We can make two approximations to generate a more convenient relationship between the microscopic and macroscopic rates. Firstly, we assume that
\begin{equation}
x_i \approx \Omega\phi_i. \label{eq:small_noise_approx}
\end{equation}
This is a small noise approximation, since it is often assumed that $x_i = \Omega \phi_i + \Omega^{1/2} \xi_i$, where $\xi_i$ is a noise term \citep{Van92}. If $\xi_i$ is small then $x_i \approx \Omega\phi_i$ is a valid approximation. Secondly, we assume that
\begin{equation}
x_i(x_i-1)\dots (x_i-s_{ij}+1) \approx x_i^{s_{ij}}. \label{eq:large_copy_approx}
\end{equation} 
This is a large copy number approximation: in the case of e.g.\ a bimolecular reaction ($2X_i\rightarrow *$) with $s_{ij}=2$, the approximation is of the form $x_i(x_i-1)\approx x_i^2$ or $x_i \approx x_i-1$. By applying Eq.~\eqref{eq:large_copy_approx} to the factor of $x_i!$ in Eq.~\eqref{eq:exact_rc_reln}, the factor of $(x_i-s_{ij})!$ cancels from the left-hand side. Simplifying using Eq.~\eqref{eq:small_noise_approx}, $\phi_i^{s_{ij}}$ cancels from both sides and we arrive at the important relationship 
\begin{equation}
\tilde{k}_j \approx \hat{k}_j \prod_{i=1}^N \frac{1}{s_{ij}!}. \label{eq:rc_reln}
\end{equation}
With Eq.~\eqref{Eq:RE}, Eq.~\eqref{eq:macro_rate_fn} and Eq.~\eqref{eq:rc_reln} one may therefore write down a set of ODEs for an arbitrary chemical reaction network, with constant reaction rates, in terms of the microscopic rates $\hat{k}_j$. This equation highlights that for reactions with $s_{ij}\geq 2$, $\tilde{k}_j \neq \hat{k}_j$, as is the case for bimolecular reactions of the form $2 X_i \rightarrow *$ (see Eq.~\eqref{eq:non-lin-fus} and Eq.~\eqref{eq:si_fus1}).  

Importantly, if the microscopic rate function is a function of state then $\hat{k}=\hat{k}(\mathbf{x})$ and $\tilde{k}=\tilde{k}(\boldsymbol{\phi})\approx\tilde{k}(\mathbf{x}/\Omega)$. In this case, Eq.~\eqref{eq:rc_reln} still applies since the above argument assumed nothing about the particular forms of $\hat{k}$ and $\tilde{k}$. However, additional factors of $\Omega^{-1}$ are induced by applying Eq.~\eqref{eq:small_noise_approx}, which may carry through to the individual parameters of $\tilde{k}(\boldsymbol{\phi})$. A demonstration of this is given in the following section.

\section*{Deriving an ODE description of the mitochondrial network system}\label{SI:ODE_net_deriv}

In this section we show how to derive an ODE description of the network system described in Eq.~\eqref{eq:non-lin-fus}-Eq.~\eqref{eq:fus-cross-3} in the Main Text. In accordance with the notation in the previous section, we will redefine all of the rates in Eq.~\eqref{eq:non-lin-fus}-Eq.~\eqref{eq:lin_feedback_ctrl} with a hat notation ($\hat{a}$, for a general rate parameter $a$), to reflect that these are stochastic rates. Deterministic rates will be denoted with a tilde ($\tilde{a}$). Our aim will be to write a set of ODEs in terms of the stochastic rates, $\hat{a}$, for which we are able to estimate values.

We will begin by considering the fusion network equations Eq.~\eqref{eq:non-lin-fus} and Eq.~\eqref{eq:lin-fus}. For clarity, we rewrite Eq.~\eqref{eq:non-lin-fus} to allow the reaction to proceed with some arbitrary rate $\hat{\rho}$:
\begin{equation}
X_S + X_S \xrightarrow{\hat{\rho}} X_F + X_F, \label{eq:non-lin-fus_rewrite}
\end{equation}
where $X$ denotes either a wild-type ($W$) or mutant ($M$). We will subsequently fix $\hat{\rho}$ to the rate of all other fusion reactions $\hat{\gamma}$. We do this because Eq.~\eqref{eq:non-lin-fus} is a bimolecular reaction involving one species: a fundamentally different reaction to bimolecular reactions involving two species, as we will now see.

Since $\hat{\rho}, \hat{\gamma}=$ const, we may use Eq.~\eqref{eq:rc_reln}, resulting in the deterministic rates
\begin{eqnarray}
\tilde{\rho} = \frac{\hat{\rho}}{2}\\
\tilde{\gamma} = \hat{\gamma}
\end{eqnarray}
for Eq.~\eqref{eq:non-lin-fus_rewrite} and Eq.~\eqref{eq:lin-fus} respectively. If we then enforce the microscopic rates to be equal for both of these fusion reactions, i.e.\ $\hat{\rho} = \hat{\gamma}$, then $\tilde{\rho} = \hat{\gamma}/2$. All other fusion reactions have $\tilde{\gamma} = \hat{\gamma}$ by application of Eq.~\eqref{eq:rc_reln}. Application of Eq.~\eqref{eq:rc_reln} to the fission reaction in Eq.~\eqref{eq:fis} shows that $\hat{\beta}=\tilde{\beta}$.

For Eq.~\eqref{eq:birth}, we have chosen a $\hat{\lambda}$ which is not a constant, but a function of the copy numbers of the chemical species ($\hat{\lambda} = \hat{\lambda}(\mathbf{x})$ where $\mathbf{x}=(w_s,w_f,m_s,m_f)$, see Eq.~\eqref{eq:lin_feedback_ctrl}). As pointed out in the previous section, care must be taken in writing down the deterministic analogue of $\hat{\lambda}$. Applying Eq.~\eqref{eq:rc_reln}, we have
\begin{equation}
\hat{\mu} + \hat{b}(\hat{\kappa} - (w_s + w_f + \hat{\delta} m_s + \hat{\delta} m_f ))   = \tilde{\mu} + \tilde{b}(\tilde{\kappa} - (\phi_{w_s} + \phi_{w_f} + \tilde{\delta} \phi_{m_s} + \tilde{\delta} \phi_{m_f} )) .
\end{equation}
Applying Eq.~\eqref{eq:small_noise_approx} and equating individual terms, we arrive at
\begin{eqnarray}
\hat{\mu} &=& \tilde{\mu}\\
\hat{b} w_s = \tilde{b}\frac{w_s}{\Omega} &\implies& \hat{b} = \tilde{b}/\Omega\\
\hat{b}\hat{k} = \tilde{b}\tilde{k} &\implies& \hat{k} = \tilde{k} \Omega\\
\hat{b}\hat{\delta}m_s = \tilde{b}\tilde{\delta}\phi_{m_s} &\implies & \hat{\delta} = \tilde{\delta}.
\end{eqnarray}
In this study, we let $\Omega = 1$ so the above 4 parameters are identical to their deterministic counterparts. Hence, by application of Eq.~\eqref{Eq:RE}, we arrive at the following set of ODEs
\begin{align}
\begin{split}
\deriv{\phi_{w_s}}{t} = - 2 \cdot \frac{\hat{\gamma}}{2}  \phi_{w_s}^2 - \hat{\gamma}  \phi_{w_s} \\ \phi_{w_f} + \hat{\beta}  \phi_{w_f} - (\hat{\mu} + \hat{b}  (\hat{\kappa} - (\phi_{w_s} + \phi_{w_f}\\ + \hat{\delta}  \phi_{m_s} +  \hat{\delta}  \phi_{m_f})))  \phi_{w_s} - \hat{\mu}  \phi_{w_s} -\\ \hat{\gamma}  \phi_{m_f}  \phi_{w_s} - \hat{\gamma}  \phi_{w_s}  \phi_{m_s} \label{eq:phi_ws}
\end{split}
\end{align}
\begin{align}
\begin{split}
\deriv{\phi_{m_s}}{t} = - 2 \cdot \frac{\hat{\gamma}}{2}  \phi_{m_s}^2 - \hat{\gamma}  \phi_{m_s} \\ \phi_{m_f} + \hat{\beta}  \phi_{m_f} - (\hat{\mu} + \hat{b}  (\hat{\kappa} - (\phi_{w_s} + \phi_{w_f}\\ + \hat{\delta}  \phi_{m_s} +  \hat{\delta}  \phi_{m_f})))  \phi_{m_s} - \hat{\mu}  \phi_{m_s} -\\ \hat{\gamma}  \phi_{w_f}  \phi_{m_s} - \hat{\gamma}  \phi_{w_s}  \phi_{m_s} \label{eq:phi_ms}
\end{split}
\end{align}
\begin{align}
\begin{split}
\deriv{\phi_{w_f}}{t} = 2 \cdot \frac{\hat{\gamma}}{2}  \phi_{w_s}^2 + \hat{\gamma}  \phi_{w_s} \\ \phi_{w_f} - \hat{\beta}  \phi_{w_f} + (\hat{\mu} + \hat{b}  (\hat{\kappa} - (\phi_{w_s} + \phi_{w_f}\\ + \hat{\delta}  \phi_{m_s} +  \hat{\delta}  \phi_{m_f})))  (2\phi_{w_s} + \phi_{w_f}) +\\ \hat{\gamma}  \phi_{m_f}  \phi_{w_s} + \hat{\gamma}  \phi_{w_s}  \phi_{m_s} \label{eq:phi_wf}
\end{split}
\end{align}
\begin{align}
\begin{split}
\deriv{\phi_{m_f}}{t} = 2 \cdot \frac{\hat{\gamma}}{2}  \phi_{m_s}^2 + \hat{\gamma}  \phi_{m_s} \\ \phi_{m_f} - \hat{\beta}  \phi_{m_f} + (\hat{\mu} + \hat{b}  (\hat{\kappa} - (\phi_{w_s} + \phi_{w_f}\\ + \hat{\delta}  \phi_{m_s} +  \hat{\delta}  \phi_{m_f})))  (2\phi_{m_s} + \phi_{m_f}) +\\ \hat{\gamma}  \phi_{w_f}  \phi_{m_s} + \hat{\gamma}  \phi_{w_s}  \phi_{m_s}. \label{eq:phi_mf}
\end{split}
\end{align}

The steady state solution of this system of ODEs may be calculated, but its form is complex. For notational simplicity, we will drop the hat notation. Defining
\begin{align}
\begin{split}
x_1 = (b^2 (\beta ^2+2 \beta  (\gamma  \kappa +3 \mu )+\gamma ^2 \kappa ^2+\mu ^2+2 \gamma  \mu \\ (\kappa +2 (\delta -1) \phi_{m_s}))+2 b \gamma  \mu  (-\beta +\mu +\gamma  (\kappa -2 \delta \\ \phi_{m_s}+2 \phi_{m_s}))+\gamma ^2 \mu ^2)^{1/2}
\end{split}
\end{align}
the non-trivial, physically-realizable, component of the steady state may be parametrized in terms of $\phi_{m_s}$ and written as
\begin{align}
\begin{split}
\phi_{w_s} =  -(\beta  b^2 \kappa +b^2 \gamma  \kappa ^2+b^2 \kappa  \mu +2 b^2 \delta  \mu \\ \phi_{m_s}+\beta ^2 b+\beta  b \gamma  \kappa +5 \beta  b \mu +3 b \gamma  \kappa  \mu -\beta  \gamma  \\ \mu +2 b \mu ^2-2 b \gamma  \delta  \mu  \phi_{m_s}-2 b \gamma  \mu  \phi_{m_s}-x_1 (b \kappa +\\\beta +2 \mu )+2 \gamma  \mu ^2+2 \gamma ^2 \mu  \phi_{m_s})/(2 \mu  (b-\gamma )^2) \label{eq:det_ss_ws}
\end{split}
\end{align}
\begin{align}
\begin{split}
\phi_{w_f} = (b^3 (\beta ^2 \kappa +\beta  (\gamma  \kappa  (\kappa +(\delta -1) \phi_{m_s})+\mu  (3\\ \kappa +\delta  \phi_{m_s}))+\phi_{m_s} (\gamma ^2 (\delta -1) \kappa ^2-\delta  \mu ^2+\gamma  \mu  \\ ((2 \delta -3) \kappa +2 (\delta -1) \delta  \phi_{m_s})))+b^2 (\beta ^3+\beta ^2 (6 \mu +\gamma \\ (\kappa +(\delta -1) \phi_{m_s}))+\beta  (5 \mu ^2+\gamma ^2 \delta  \kappa  \phi_{m_s}+\gamma ^2 \\ \kappa  (-\phi_{m_s})+6 \gamma  \delta  \mu  \phi_{m_s}-8 \gamma  \mu  \phi_{m_s}-\kappa  x_1)+\\\phi_{m_s} (2 \gamma ^2 \mu  (\delta  \kappa +\delta ^2 (-\phi_{m_s})+\phi_{m_s})+\gamma  ((6 \delta \\ -5) \mu ^2-(\delta -1) \kappa  x_1)-\delta  \mu  x_1))-b (\beta  (\gamma ^2 \mu  (-(\kappa +(4-3 \\ \delta ) \phi_{m_s}))+\gamma  (2 \mu ^2+(\delta -1) \phi_{m_s} x_1)+3 \mu  x_1)+\gamma  \mu \\ \phi_{m_s} (\gamma  (\delta -2) \mu +\gamma ^2 (\kappa -2 (\delta -1) \phi_{m_s})+(\delta -3) x_1)+ \\\ \beta ^2 (2 \gamma  \mu +x_1))+\gamma  \mu  (\beta -\gamma  \phi_{m_s}) (\gamma  \mu +x_1))/(2 b \\ \mu  (b-\gamma )^2 (\beta +\gamma  (\delta -1) \phi_{m_s})) \label{eq:det_ss_wf}
\end{split}
\end{align}
\begin{align}
\begin{split}
\phi_{m_f} = (\phi_{m_s} (-b \beta +b \gamma  \kappa +b \mu -2 b \gamma  \delta  \phi_{m_s}+2 b \\ \gamma  \phi_{m_s}+\gamma  \mu +x_1))/(2 b (\beta +\gamma  (\delta -1) \phi_{m_s})) \label{eq:det_ss_mf}.
\end{split}
\end{align}
Since the steady state is parametrized by $ \phi_{m_s}$, the steady state is therefore a line.

\section*{Proof of heteroplasmy relation for linear feedback control}\label{SI:proof_vh_lfc}

In this section we show that Eq.~\eqref{eq:ansatz} holds for the system described by Eq.~\eqref{eq:non-lin-fus}-Eq.~\eqref{eq:fus-cross-3} given the replication rate in Eq.~\eqref{eq:lin_feedback_ctrl} using the Kramers-Moyal expansion under conditions of large copy number and fast network churn (to be defined below); the approach used here is similar to \cite{Constable16}. Consonant with the self-contained objectives of STAR methods, we draw together elements from the literature to provide a coherent derivation; we therefore hope that the following exposition may provide clarity for a wider audience.

\paragraph*{Kramers-Moyal expansion of the chemical master equation for large copy numbers}

Customarily, the Kramers-Moyal expansion is formed using a continuous-space notation \citep{Gardiner85}, so we will initially proceed in this way. Following the treatment by \cite{Gardiner85}, we begin by re-writing the chemical master equation Eq.~\eqref{Eq:CME} (CME) as
\begin{equation}
\pderiv{P(\x,t)}{t}= \int_{-\infty}^{\infty}\d{}\x{}'\left[T(\x{}|\x{}')P(\x{}',t)-T(\x{}'|\x{})P(\x{},t)\right]\label{eq:master_eqn_continuous_omega}
\end{equation}
where we have set $\Omega = 1$. $T(\x{}|\x{}')$ is the transition rate from state $\x{}'\rightarrow \x{}$, and the dependence upon the initial condition has been suppressed for notational convenience. We now proceed by expanding the CME. The multivariate Kramers-Moyal expansion may be written as 
\begin{equation}
\pderiv{P(\x,t)}{t} \approx \int_{-\infty}^{\infty}\left(-\nabla \left(T(\x'|\x)P(\x)\right)^T \cdot (\x'-\x) + \frac{1}{2} (\x'-\x)^T \cdot \mathbf{H} \cdot (\x'-\x) \right)\d{}\x' \label{eq:KM_mv}
\end{equation}
where $\mathbf{H}(\x)$ is the Hessian matrix of $T(\x'|\x)P(\x)$
\begin{equation}
\mathbf{H}\defeq\begin{pmatrix}
\frac{\partial^2}{\partial x_1^2}  & \dots & \frac{\partial^2}{\partial x_1 \partial x_N}  \\
\vdots & & \vdots \\
\frac{\partial^2}{\partial x_N \partial x_1}  & \dots & \frac{\partial^2}{\partial x_N^2} 
\end{pmatrix} T(\x'|\x)P(\x) \label{eq:Hessian}
\end{equation}
(see \citep{Gardiner85} for a proof of this in the univariate case). 

A transition to each possible neighbouring state $\x'$ corresponds to some reaction $j$ which moves the state from $\x \rightarrow \x'$. Since we know the influence of each reaction on state $\x$ through the constant stoichiometry matrix $S_{ij}$, and that the propensity of a reaction does not depend upon $\x'$ itself (see Eq.~\eqref{eq:micro_rate_fn}), we may transition from a notation involving $\x$ and $\x'$ into a notation involving $\x$ and $j$. We may therefore define $T_j(\x) \defeq T(\x'|\x) \equiv \hat{f}_j(\x)$ (see Eq.~\eqref{eq:micro_rate_fn}), and let $\mathbf{H}(\x) \rightarrow \mathbf{H}_j(\x)$.

We now make a large copy number assumption in order to simplify $T_j(\x{})$. To take a large copy number limit, we assume that $x_i!\approx x_i^{s_{ij}} (x_i - s_{ij})!$ resulting in
\begin{equation}
T_j(\x) \approx \hat{k}_j \prod_{i=1}^N \frac{x_i^{s_{ij}}}{s_{ij}!}. \label{eq:trans_rate_large_copy}
\end{equation}
This approximation is exact when $s_{ij}=0,1$, but inexact when $s_{ij}\geq 2$. For example, if we consider the second-order bimolecular reaction in Eq.~\eqref{eq:non-lin-fus}, Eq.~\eqref{eq:trans_rate_large_copy} is equivalent to assuming $w_s^2 \approx w_s (w_s - 1)$; consequently, a factor of $1/(s_{ij}!) = 1/2$ arises in $T_j(\x)$ as a combinatorial factor from stochastic considerations.

\paragraph*{Fokker-Planck equation for chemical reaction networks}

We now wish to re-write Eq.~\eqref{eq:KM_mv} as a Fokker-Planck equation. Since the integral in Eq.~\eqref{eq:KM_mv} is over $\x'$, and every $\x'$ corresponds to a reaction $j$, we may interpret the integral in Eq.~\eqref{eq:KM_mv} as a sum over all reactions, i.e.\ $\int \d{}\x' \rightarrow \sum_{j=1}^R$. Hence, for the $j$\textsuperscript{th} reaction, $[(\x'-\x)]_i = S_{ij}$. With these observations, we may write the first integral of Eq.~\eqref{eq:KM_mv} as
\begin{align}
\int_{-\infty}^{\infty}-\nabla (T(\x'|\x)P(\x))^T \cdot (\x'-\x) \d{}\x' &= \int_{-\infty}^{\infty}-\nabla (T_j(\x)P(\x,t))^T \cdot (\x'-\x) \d{}\x' \nonumber \\
&= -\sum_{j=1}^R \sum_{i=1}^N \pderiv{}{x_i}( T_j(\x) P(\x,t))  S_{ij} \nonumber \\
&= - \sum_{i=1}^N \pderiv{}{x_i} A_i P(\x,t)
\end{align}
where
\begin{equation}
\mathbf{A} \defeq \mathbf{S}\cdot \mathbf{T}.
\end{equation}
$\mathbf{A}$ is a vector of length $N$, $[\mathbf{S}]_{ij} \defeq r_{ij} - s_{ij}$ is the $N\times R$ stoichiometry matrix Eq.~\eqref{eq:gen_chem_sys}, and $\mathbf{T}$ is the vector of transition rates, of length $R$ (for which we have taken a large copy number approximation in Eq.~\eqref{eq:trans_rate_large_copy}). To re-write the second integral of Eq.~\eqref{eq:KM_mv}, we write an element of the Hessian $\mathbf{H}_j$ in Eq.~\eqref{eq:Hessian} as
\begin{equation}
H_{jlm} = \frac{\partial^2}{\partial x_l \partial x_m} T_j(\x) P(\x,t)
\end{equation}
where $j=1,\dots,R$ and $l,m=1,\dots,N$. Thus, we may write
\begin{align}
\int_{-\infty}^{\infty} \frac{1}{2} (\x'-\x)^T \cdot \mathbf{H}_{j} \cdot (\x'-\x) \d{}\x' &= \frac{1}{2} \sum_{j=1}^R \sum_{l=1}^N \sum_{m=1}^N S_{lj} H_{jlm} S_{mj} \nonumber \\
&= \frac{1}{2} \sum_{j=1}^R \sum_{l=1}^N \sum_{m=1}^N S_{lj} \frac{\partial^2}{\partial x_l \partial x_m} T_j P(\x,t) S_{mj} \nonumber \\
&= \frac{1}{2} \sum_{l=1}^N \sum_{m=1}^N  \frac{\partial^2}{\partial x_l \partial x_m} \left( \sum_{j=1}^R S_{lj} T_j S_{mj} \right) P(\x,t) \nonumber \\
&= \frac{1}{2} \sum_{i,m=1}^N   \frac{\partial^2}{\partial x_i \partial x_m} B_{im} P(\x,t)
\end{align}
where
\begin{equation}
\mathbf{B} \defeq \mathbf{S} \cdot \Diag(\mathbf{T}) \cdot \mathbf{S}^T. \label{eq:diff_KM_fpe}
\end{equation}
$\mathbf{B}$ is an $N\times N$ matrix, and $\Diag(\mathbf{Y})$ is a diagonal matrix whose main diagonal is the vector $\mathbf{Y}$. We may therefore re-write Eq.~\eqref{eq:KM_mv} as a Fokker-Planck equation for the state vector $\x$ of the form
\begin{equation}
\pderiv{P(\x,t)}{t} \approx - \sum_{i=1}^N \pderiv{}{x_i} [A_i(\x)P(\x,t)]+\frac{1}{2}\sum_{i,m=1}^N \pdd{}{x_i}{x_m}[B_{im}(\x)P(\x)]. \label{eq:FPE}
\end{equation}

\paragraph*{Fokker-Planck equation for an arbitrary function of state}

We now wish to make a change of variables in Eq.~\eqref{eq:FPE} to write down a Fokker-Planck equation for an arbitrary scalar function of state $\x$ (which we will later set to be heteroplasmy). To do this, we wish to make use of It\^{o}'s formula, which allows a change of variables for an SDE. In general, the Fokker-Planck equation in Eq.~\eqref{eq:FPE} is equivalent \citep{Jacobs10} to the following It\^{o} stochastic differential equation (SDE)
\begin{equation}
\d{} \x = \mathbf{A} \d{} t + \mathbf{G} \d{} \mathbf{W} \label{eq:SDE}
\end{equation}
where $\mathbf{G}\mathbf{G}^T\equiv\mathbf{B}$ (where $\mathbf{G}$ is an $N \times R$ matrix)  and $\d{} \mathbf{W}$ is a vector of independent Wiener increments of length $R$, and a Wiener increment $\d W$ satisfies 
\begin{equation}
\int_0^t \d W \defeq W(t),\ P(W,t) \equiv \frac{1}{\sqrt{2 \pi t}} e^{-W^2/(2t)}.
\end{equation}
It\^{o}'s formula states that, for an arbitrary function $h(\x,t)$ where $\x$ satisfies Eq.~\eqref{eq:SDE}, we may write the following SDE
\begin{equation}
\d h(\x,t) = \left\{ \pderiv{h}{t} + \left(\nabla h\right)^T \mathbf{A} + \frac{1}{2} \Tr \left[ \mathbf{G}^T \mathbf{H}_h(\x) \mathbf{G} \right]   \right\} \d t + (\nabla h)^T \mathbf{G} \d{} \mathbf{W}, \label{eq:ito_formula}
\end{equation}
where $\mathbf{H}_h(\x)$ is the Hessian matrix of $h(\x,t)$ (see Eq.~\eqref{eq:Hessian}, where $T(\x'|\x) P(\x)$ should be replaced with $h(\x,t)$). Given the form of $\mathbf{B}$ in Eq.~\eqref{eq:diff_KM_fpe} we let
\begin{equation}
\mathbf{G} = \mathbf{S}\cdot \Diag(\sqrt{\mathbf{T}}),
\end{equation}
which satisfies $\mathbf{G}\mathbf{G}^T\equiv\mathbf{B}$.

For convenience, we may also perform the transformation purely at the level of Fokker-Planck equations. Let $h(\x,t)$ satisfy the general Fokker-Planck equation
\begin{equation}
\pderiv{P(h,t)}{t} = - \pderiv{}{h} [\tilde{A}(h,t)P(h,t)]+\frac{1}{2} \frac{\partial^2 }{\partial h^2} [\tilde{B}(h,t) P(h,t) ] \label{eq:FPE_tfm}
\end{equation}
for scalar functions $\tilde{A}(h,t)$ and $\tilde{B}(h,t)$. Using the cyclic property of the trace in Eq.~\eqref{eq:ito_formula}, we may identify 
\begin{equation}
\tilde{A} =  \pderiv{h}{t} + \left(\nabla h\right)^T \mathbf{A} + \frac{1}{2} \Tr \left[ \mathbf{B} H_h(\x) \right] \label{eq:FPE_tfm_A}
\end{equation}
where $\Tr$ is the trace operator. Also, from Eq.~\eqref{eq:ito_formula},
\begin{equation}
\tilde{B} = [(\nabla h)^T \mathbf{G}] [(\nabla h)^T \mathbf{G}]^T = (\nabla h)^T \mathbf{B} (\nabla h). \label{eq:FPE_tfm_B}
\end{equation}
Hence, using Eq.~\eqref{eq:FPE_tfm}, Eq.~\eqref{eq:FPE_tfm_A} and Eq.~\eqref{eq:FPE_tfm_B}, we may write down a Fokker-Planck equation for an arbitrary function of state in terms of $\mathbf{A}$ and $\mathbf{B}$.

\paragraph*{An SDE for heteroplasmy forced onto the steady state line in the high-churn limit}

It has been demonstrated that SDE descriptions of stochastic systems which possess a globally-attracting line of steady states may be formed in the long-time limit by forcing the state variables onto the steady state line \citep{Constable16,Parsons17}. Such descriptions may be formed in terms of a parameter which traces out the position on the steady state line, hence reducing a high-dimensional problem into a single dimension \citep{Constable16,Parsons17}. In our case, heteroplasmy is a suitable parameter to trace out the position on the steady state line. We seek to use similar reasoning to verify Eq.~\eqref{eq:ansatz}. In what follows, we will assume that $\x(t=0) = \xss$, where $\xss$ is the state which is the solution of $\mathbf{A}=\mathbf{0}$ (which is equivalent to finding the steady state solution of the deterministic rate equation in Eq.~\eqref{Eq:RE} due to our assumption of large copy numbers and $\Omega=1$),  so that we may neglect any deterministic transient dynamics.

Inspection of the steady state of the ODE description of our system reveals that the set of steady state solutions forms a line (see Eqs.~\eqref{eq:det_ss_ws}--\eqref{eq:det_ss_mf}). Inspection of the steady state solution reveals that the steady state depends on the fusion ($\gamma$) and fission ($\beta$) rates. Mitochondrial network dynamics occur on a much faster timescale than the replication and degradation of mtDNA: the former occurring on the timescale of minutes \citep{Twig08} whereas the latter is hours or days \citep{Johnston16}. We seek to use this separation of timescales to arrive at a simple form for $\mathbb{V}(h)$. We redefine the fusion and fission rates such that
\begin{eqnarray}
\gamma &\rightarrow& M \gamma \nonumber \\
\beta &\rightarrow& M \beta
\end{eqnarray}
where $M$ is a constant which determines the magnitude of the fusion and fission rates, which we call the ``network churn''.

We now wish to use heteroplasmy
\begin{equation}
h(\x,t)= h(\x) \defeq (m_s+m_f)/(w_s+w_f+m_s+m_f), 
\end{equation}
as our choice for the function of state in the Fokker-Planck equation in Eq.~\eqref{eq:FPE_tfm}.  We will first compute the diffusion term $\tilde{B}$ for heteroplasmy using Eq.~\eqref{eq:FPE_tfm_B}. If we constrain the state $\x$ to be forced onto the steady state line $\xss$ (as per \citep{Constable16,Parsons17}) in the high-churn limit, then upon defining 
\begin{equation}
\theta \defeq \sqrt{b^2 (\beta +\gamma  \kappa )^2-2 b \gamma  \mu  \left(\beta -\gamma  \kappa +2 \gamma  (\delta -1) m_s\right)+\gamma ^2 \mu ^2}
\end{equation}
we have
\begin{align}
\begin{split}
\lim_{M\rightarrow \infty}\left(\left. \tilde{B}\right|_{\x = \xss}\right) = (16 b^2 \gamma ^2 \mu \\  m_s (\beta +\gamma  (\delta -1) m_s){}^2 (b^2 (\beta ^3+2 \beta ^2 \gamma \\  (\delta -1) m_s+\beta  \gamma ^2 (\kappa ^2+(1-2 \delta ) \kappa  m_s+(\delta -1) (2 \delta -1) \\  m_s^2)+\gamma ^3 \kappa  m_s ((\delta -1) m_s-\kappa ))+b (-\beta ^2 \\  (2 \gamma  \mu +\theta )+\beta  \gamma  (\kappa  (2 \gamma  \mu +\theta )+m_s (\gamma  (5-4 \\  \delta ) \mu -2 (\delta -1) \theta ))+\gamma ^2 m_s ((\delta -1) m_s (3 \gamma  \mu \\ +\theta )-\kappa  (2 \gamma  \mu +\theta )))+\gamma  \mu  (\gamma  \mu +\theta ) \\  (\beta -\gamma  m_s)))/(\beta ^3 (b (\gamma  (\kappa -2 \delta  m_s \\ +2 m_s)-\beta )+\gamma  \mu +\theta ){}^4). \label{eq:diff_lin_ctrl_het}
\end{split}
\end{align}

Eq.~\eqref{eq:diff_lin_ctrl_het} is difficult to understand. In order to perform further simplification, we make an ansatz for the form of $\tilde{B}$ ($\tilde{B}_{\text{An}}$) and seek to determine whether our ansatz is equivalent to the derived form of $\tilde{B}$ under the constraints defined on the left-hand side of Eq.~\eqref{eq:diff_lin_ctrl_het}. Our ansatz takes the form
\begin{equation}
\tilde{B}_{\text{An}} \defeq \lim_{M\rightarrow \infty}\left(\left. \frac{2 \mu h (1-h)}{n(\x)}\cdot f_s(\x) \right|_{\x = \xss}\right) \label{eq:ansatz_B}
\end{equation}
where $f_s(\x) \defeq (w_s+m_s)/(w_s+w_f+m_s+m_f)$ and $n(\x) \defeq w_s + w_f + m_s + m_f$. Notice that this ansatz is more general than Eq.~\eqref{eq:diff_lin_ctrl_het}, since it has no explicit dependence upon the parameters of the control law assumed in Eq.~\eqref{eq:lin_feedback_ctrl}, and only explicitly depends upon functions of state $\x$.

Upon substituting the steady state $\xss$ into the ansatz in Eq.~\eqref{eq:ansatz_B} and taking the high-churn limit, we find that
\begin{align}
\begin{split}
\tilde{B}_{\text{An}} = -(m_s (\beta +\gamma  (\delta -1) m_s) \\ (b (\beta +\gamma  \kappa )-\gamma  \mu +\theta ) \\ (b (\beta +\gamma  \kappa )+\gamma  \mu -\theta ) (b (m_s (2 \beta  \delta -\beta \\ +\gamma  \kappa )-2 \beta  \kappa )+m_s (\theta -\gamma  \mu )))/(2 b \beta ^3 (\kappa -\delta  \\ m_s+m_s)^2 (b (\gamma  (\kappa -2 \delta  m_s+2 m_s)-\beta )+\gamma  \mu +\theta )). \label{eq:diff_lin_ctrl_het_an}
\end{split}
\end{align}
After some algebra (see GitHub repository for Mathematica notebook), it can be shown that Eq.~\eqref{eq:diff_lin_ctrl_het} and Eq.~\eqref{eq:diff_lin_ctrl_het_an} are equivalent, i.e.\ 
\begin{equation}
\tilde{B}_{\text{An}} \equiv \lim_{M\rightarrow \infty}\left(\left. \tilde{B}\right|_{\x = \xss}\right).
\end{equation} 
As such, we may use $\tilde{B}$ and $\tilde{B}_{\text{An}}$ interchangeably in the limit of high network churn. Furthermore, it can be shown after some algebra that the drift of heteroplasmy when forced onto the steady state line is 0, i.e.\
\begin{equation}
\left. \tilde{A}\right|_{\x = \xss} \equiv 0. \label{eq:drift_ss_0}
\end{equation} 
A similar result is shown in \citep{Constable16} ((Equation S59) therein). Substituting $h(\x,t)=h$, $\tilde{A}$ and $\tilde{B}$ into the Fokker-Planck equation for an arbitrary function of state Eq.~\eqref{eq:FPE_tfm}, we have
\begin{equation}
\pderiv{P(h,t)}{t} = \frac{1}{2} \frac{\partial^2 }{\partial h^2} \left[\left.\left(\frac{2 \mu h (1-h)}{n(\x)}\cdot f_s(\x)\right)\right|_{\x=\xss(h)} P(h,t) \right]
\end{equation}
which is equivalent to the following SDE for heteroplasmy
\begin{equation}
\d h = \left.\sqrt{\frac{2 \mu h (1-h) f_s(\x)}{n(\x)}}\right|_{\x=\xss(h)} \d W \label{eq:SDE_network_sys}
\end{equation}
in the limit of large network churn, large copy numbers, and a second-order truncation of the Kramers-Moyal expansion. Although the state has been forced onto the steady state, stochastic fluctuations mean that trajectories may move along the line of steady states, so the diffusion coefficient is not constant in general. We may calculate the new value of $\xss(h)$ for every displacement due to Wiener noise in $h$,  and substitute into $f_s(\x)$ and $n(\x)$ to determine the diffusion coefficient at the next time step.

However, for sufficiently short times, and large copy numbers (i.e.\ low diffusivity of $h$), we may assume that the diffusion coefficient in Eq.~\eqref{eq:SDE_network_sys} may be approximated as constant. Since the general solution of the SDE
\begin{equation}
\d y = \sqrt{B} \d W
\end{equation}
for $B=$ const is 
\begin{equation}
y \sim \mathcal{N}(y|y_0,Bt)
\end{equation} 
where $\mathcal{N}(y|y_0,\sigma^2)$ is a Gaussian distribution on $y$ with mean $y_0$ and variance $\sigma^2$, and $y_0 = y(t=0)$. Since we have assumed that the state is initialised at $\x(t=0)=\xss$, there are no deterministic transient dynamics, so we may write
\begin{equation}
\mathbb{V}(h)\approx \left. \frac{2 \mu t}{n(\x)} h(\x) (1-h(\x)) f_s(\x) \right|_{\x=\xss}, \label{eq:ansatz_derived}
\end{equation}
where $\mathbb{V}$ returns variance of a random variable. In this equation, we take $\x=\xss=$ const, since we have assumed a low-diffusion limit. We observe that this equation is of precisely the same form as (Equation 12) of \cite{Johnston16}, except with an additional proportionality factor of $f_s$ induced by the inclusion of a mitochondrial network.

\section*{Heteroplasmy variance relations for alternative model structures and modes of genetic mtDNA control}\label{SI:Vh_alt_struct}

Here we explore the implications of alternative model structures upon Eq.~\eqref{eq:SDE_network_sys}. Firstly, we may consider replacing Eq.~\eqref{eq:birth} with
\begin{equation}
X_S \xrightarrow{\lambda} X_S + X_S. \label{eq:s_birth_s}
\end{equation}
This corresponds to the case where replication coincides with fission, see \citep{Lewis16}. Repeating the calculation in the previous section also results in Eq.~\eqref{eq:SDE_network_sys}, so the result is robust to the particular choice of mtDNA replication reaction (see GitHub repository for Mathematica notebook).

Secondly, we may explore the impact of allowing non-zero mtDNA degradation of fused species. This could correspond to autophagy-independent degradation of mtDNA, for example via the exonuclease activity of POLG \citep{Medeiros18}. To encode this, we may add the following additional reaction
\begin{equation}
X_F \xrightarrow{\xi\mu} \emptyset \label{eq:fused_deg_xi}
\end{equation}
where $0 \leq \xi \leq 1$. We were not able to make analogous analytical progress in this instance. However, numerical investigation (Figure~\ref{Fig:stoch_extras}E) revealed that the following ansatz was able to predict heteroplasmy variance dynamics
\begin{equation}
\mathbb{V}(h)\approx \left. \frac{2 \mu t}{n(\x)} h(\x) (1-h(\x)) (f_s(\x) + \xi (1-f_s(\x))) \right|_{\x=\xss}. \label{eq:ansatz_xi}
\end{equation}
In other words, allowing degradation of fused species results in a linear correction to our heteroplasmy variance formula in Eq.~\eqref{eq:ansatz}. If fused species are susceptible to degradation at the same rate as unfused species ($\xi=1$), then $\mathbb{V}(h)$ loses $f_s$ dependence entirely and the mitochondrial network has no influence over heteroplasmy dynamics.

\begin{figure}[h!]
\begin{center}
\includegraphics[width=0.6\columnwidth]{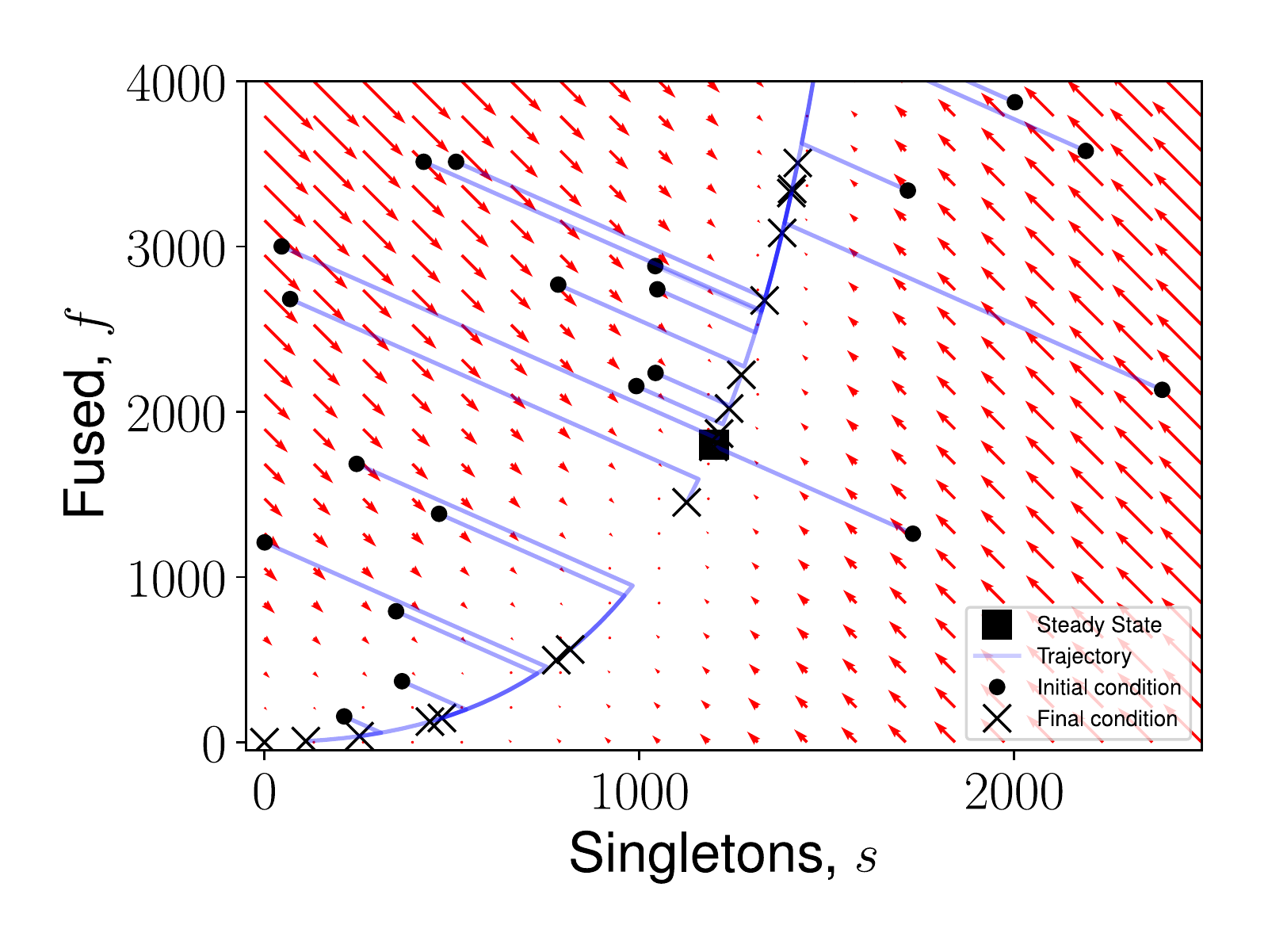}  
\end{center}
\caption{\textbf{Phase portrait for an ODE representation of a network system with constant rates}. The system displays two steady states: a trivial steady state at $s+f=0$, and a non-trivial steady state. At a steady state, both time derivatives in Eq.~\eqref{eq:si_single_genet_sdot} and Eq.~\eqref{eq:si_single_genet_fdot} vanish. Trajectories (blue) show the evolution of the system up to $t=1000$. The direction and magnitude of the derivative at points in space are shown by red arrows. Trajectories can be seen either decaying to $s=f=0$ or tending to infinity. $\gamma = 0.01656$, $\beta = 33.12$, $\rho=0.023$, $\eta=1.1$, $\alpha=1.04$. }
\label{Fig:justification}
\end{figure}

We also explored various different forms of $\lambda(\x)$ and $\mu(\x)$, which we label A-G after \citep{Johnston16}, and X-Z for several newly-considered functional forms, see Table~\ref{tab:param_sweeps} and Figure~\ref{Fig:sweeps}A-I. Control D of \citep{Johnston16} involves no feedback, which we do not explore -- see Figure~\ref{Fig:justification}, and the discussion surrounding Eq.~\eqref{eq:si_fus1}. The argument presented in the previous section requires the steady state solution of the system to be solvable, since we require the explicit form of $\xss$ in Eq.~\eqref{eq:diff_lin_ctrl_het}, Eq.~\eqref{eq:diff_lin_ctrl_het_an} and Eq.~\eqref{eq:drift_ss_0}. For controls B, C, E, F, G, Y and Z in Table \ref{tab:param_sweeps}, the steady states are solvable and similar arguments to the above can be applied (see the GitHub repository for Mathematica notebooks). Controls B, C, E, F all satisfy Eq.~\eqref{eq:SDE_network_sys}; this can be shown numerically for controls A and X. However, controls G, Y and Z satisfy 
\begin{equation}
\d h = \left.\sqrt{\frac{2 \mu h (1-h)}{n(\x)}}\right|_{\x=\xss(h)} \d W. \label{eq:SDE_network_sys_indep_net}
\end{equation}
Notably, Eq.~\eqref{eq:SDE_network_sys_indep_net} does not depend on $f_s$, unlike  Eq.~\eqref{eq:SDE_network_sys} (see GitHub repository for Mathematica notebooks). This is because control of copy number occurs in the degradation rate, rather than the replication rate, for controls G, Y and Z. A modified version of a Moran process (presented below) can provide intuition for why the diffusion rate of heteroplasmy variance depends on the network state when the population is controlled through replication, and does not depend on network state when the population is controlled through degradation.

\section*{Choice of nominal parametrization}\label{SI:nom_param}

In this section we discuss our choice of nominal parametrization for the network system in Eq.~\eqref{eq:non-lin-fus}-Eq.~\eqref{eq:fus-cross-3}, given the replication rate in Eq.~\eqref{eq:lin_feedback_ctrl}. We will first discuss our choice of network parameters. 

\cite{Cagalinec13} found that the average fission rate in cortical neurons is 0.023$\pm$0.003~fissions/mitochondria/min. Assuming that this value is representative of the fission rate in general, and converting this to units of per day, we may write the mitochondrial fission rate as $\beta = 33.12$~day$^{-1}$. 

The dimensions of $\beta$ are day$^{-1}$ and not mitochondrion$^{-1}$~day$^{-1}$. This is because if the propensity (see Eq.~\eqref{eq:propensity}, where $\Omega = 1$) of e.g. Eq.~\eqref{eq:fis} is $\hat{a}_{\text{fis},w} = \beta w_f$ then the mean time to the next event is $1/(\beta w_f)$; therefore the dimension of $\beta$ is per unit time and copy numbers are pure numbers, i.e.\ dimensionless. Similar reasoning constrains the dimension of the fusion rate, see below.

Evaluation of the fusion rate is more involved, since fusion involves two different chemical species coming together to react whereas fission may be considered as spontaneous. Furthermore, there are 7 different fusion reactions whereas there are only 2 fission reactions. For simplicity, assume that all species have a steady-state copy number of $x_i = 250$ (resulting in a total copy number of 1000, heteroplasmy of 0.5 and 50\% of mitochondria existing in the fused state). Neglecting subtleties relating to bimolecular reactions involving one species (see Eq.~\eqref{eq:rc_reln}), each fusion reaction proceeds at rate $\hat{a}_{\text{fus},j} \approx \gamma x_i^2$. Since there are 7 fusion reactions (Eq.~\eqref{eq:non-lin-fus}, Eq.~\eqref{eq:lin-fus}, Eq.~\eqref{eq:fus-cross-1}-Eq.~\eqref{eq:fus-cross-3}), the total fusion propensity is $\hat{a}_{\text{fus}} \approx 7 \gamma x_i^2$. Similarly, the total fission propensity is $\hat{a}_{\text{fis}} = \beta(w_f+m_f) = 2 \beta x_i$. Since we expect macroscopic proportions of both fused and fissioned species in many physiological settings, we may equate the fusion and fission propensities, $\hat{a}_{\text{fus}}=\hat{a}_{\text{fis}}$, and rearrange for the fusion rate $\gamma$ to yield $\gamma = 2\beta x_i/(7 x_i^2) \approx 3.8 \times 10^{-2}$~~day$^{-1}$. The orders of magnitude difference between $\beta$ and $\gamma$ stems from the observation that fusion propensity depends on the square of copy number whereas the fission propensity depends on copy number linearly.

Given the network parameters, we then explored appropriate parametrizations for the genetic parameters: the mitophagy rate ($\mu$) and the parameters of the linear feedback control ($\kappa$, $b$ and $\delta$, see Eq.~\eqref{eq:lin_feedback_ctrl}). mtDNA half-life is observed to be highly variable: in mice this can be between 10-100~days~\citep{Burgstaller14}. For consistency with another recent study investigating the relationship between network dynamics and heteroplasmy, we use an mtDNA half-life of 30~days~\citep{Tam15}.

The parameter $\delta$ in the replication feedback control (see Eq.~\eqref{eq:lin_feedback_ctrl}) may be interpreted as the ``strength of sensing of mutant mtDNA'' in the feedback control~\citep{Hoitzing17}. Assuming that fluctuations in copy number of mutants and wild-type molecules are sensed identically (as may be the case for e.g.\ non-coding mtDNA mutations) we may reasonably assume a model of $\delta=1$ as the \textit{simplest} case of a neutral mutation \ud{(although $\delta \neq 1$ still defines a neutral model, since both mutant and wild-type alleles experience the same replication and degradation rates per molecule, see Eqs.\eqref{eq:birth}--\eqref{eq:death_s})}.

We are finally left with setting the parameters $\kappa$ and $b$ in the linear feedback control Eq.~\eqref{eq:lin_feedback_ctrl}. In the absence of a network state and mutants, $\kappa$ is precisely equal to the steady state copy number, since the degradation rate equals the replication rate when $\kappa = w$. However, the presence of a network means that a subpopulation of mtDNAs (namely the fused species) are immune to death, resulting in $\kappa$ no longer being equivalent to the steady state copy number. The parameter $b$ may be interpreted as the feedback control strength, which determines the extent to which the replication rate changes given a unit change in copy number. 

Given a particular value of $b$, we may search for a $\kappa$ which gives a total steady state copy number ($n$) which is closest to some target value (e.g.\ 1000 as a typical total mtDNA copy number per cell in human fibroblasts~\citep{Kukat11}). We swept a range of different values of $b$ and found that, for values of $b$ smaller than a critical value ($b^*$), a $\kappa$ could not be found whose deterministic steady state was sufficiently close to $n=1000$. This result is intuitive because in the limit of $b \rightarrow 0$, $\lambda =$ const. From the analysis above we have shown that constant genetic rates ($\mu, \lambda$) result in unstable copy numbers, and therefore a sufficiently small value of $b$ is not expected to yield a stable non-trivial steady state solution. We chose $b \approx b^*$, and the corresponding $\kappa$, such that the steady state copy number is controlled as weakly as possible given the model structure.

\section*{Rate renormalization}\label{SI:rate_renorm}

In Eqs.~\eqref{eq:non-lin-fus}--\eqref{eq:lin_feedback_ctrl} we have neglected reactions such as 
\begin{equation}
X_F + X_F \rightarrow  X_F + X_F
\end{equation}
because they do not change the number of molecules in our state vector $\x = (w_s,w_f,m_s,m_f)$. One may ask whether neglecting such reactions means that it is necessary to renormalize the fission-fusion rates which were estimated in the preceding section. In estimating the nominal parametrization above, we began by using a literature value for the mitochondrial fission rate, and then matched the fusion rate such that the summed hazard of a fusion event approximately balanced the fission rate. This matching procedure is reasonable, since we observe a mixture of fused and fissioned mitochondria under physiological conditions: choosing a fusion rate which is vastly different results in either a hyperfused or fragmented network. We must therefore only justify the fission rate. Eq.~\eqref{eq:fis} assumes that a fission reaction always results in a singleton, and a singleton is by definition a molecule which is susceptible to mitophagy (see Eq.~\eqref{eq:death_s}). Therefore, if fission reactions always result in mitochondria containing single mtDNAs which are susceptible to mitophagy, then we expect our model to match well to true physiological rates. If, on the other hand, fission reactions between large components of the network which are too large to be degraded are common, then renormalization of $\beta$ by the fraction of fission events which result in a sufficiently small mitochondrion would be necessary, which would in turn renormalize $\gamma$ through our matching procedure. We are not aware of experimental measurements of the fraction of fission events which result in mitochondria which contain a particular number of mtDNAs. Such an experiment, combined with the distribution of mitochondrial sizes which are susceptible to mitophagy, would allow us to validate our approach. Despite this, the robustness of our results over approximately 4 orders of magnitude for the fission-fusion rate (Figure~\ref{Fig:sweeps}A-I) provides some indication that our results are likely to hold in physiological regions of parameter space.

\section*{A modified Moran process may account for the alternative forms of heteroplasmy variance dynamics under different models of genetic mtDNA control}\label{SI:mod_moran}

We sought to gain insight into why control of population size through the replication rate ($\lambda = \lambda(\x)$, $\mu =$ const) results in heteroplasmy variance depending on the fraction of unfused mitochondria (see Eq.~\eqref{eq:ansatz}), whereas control of population size through the degradation rate ($\mu = \mu(\x)$, $\lambda = $ const) results in heteroplasmy variance becoming independent of network state, where
\begin{equation}
\mathbb{V}(h)\approx \left. \frac{2 \lambda t}{n(\x)} h(\x) (1-h(\x)) \right|_{\x=\xss}. \label{eq:ansatz_indep_net}
\end{equation}
We will proceed by considering an analogous Moran process to the set of reactions presented in Eqs.~\eqref{eq:non-lin-fus}~\eqref{eq:fus-cross-3}.

First, consider a haploid biallelic Moran process consisting of wild-types and mutants, in a population of fixed size $n$. At each step in discrete time, a member of the population is chosen for birth, and another for death. Let $m_t$ denote the copies of mutants at time $t$. Then,
\begin{equation}
P(m_{t+1}=j|m_t=i)=\begin{cases}
i(n-i)/n^2& \text{if } j = i\pm 1\\
i^2/n^2 + (n-i)^2/n^2 & \text{if } j = i\\
0 & \text{otherwise.}
\end{cases} \label{eq:moran_defn}
\end{equation}
It follows that
\begin{equation}
\mathbb{E}(m_{t+1}|m_t)=m_t. 
\end{equation}
Defining $h_t\defeq m_t/n$ then from Eq.~\eqref{eq:moran_defn}
\begin{equation}
\mathbb{V}(m_{t+1}|m_t)=2 h_t (1-h_t)
\end{equation}
and therefore
\begin{equation}
\mathbb{V}\left(\left.\frac{m_{t+1}}{n}\right|m_t\right)=\mathbb{V}(h_{t+1}|m_t)=\frac{2}{n^2}h_t (1-h_t).\label{eq:h_var_moran_simple}
\end{equation}

\begin{figure}[h!]
\begin{center}
\includegraphics[width=0.95\columnwidth]{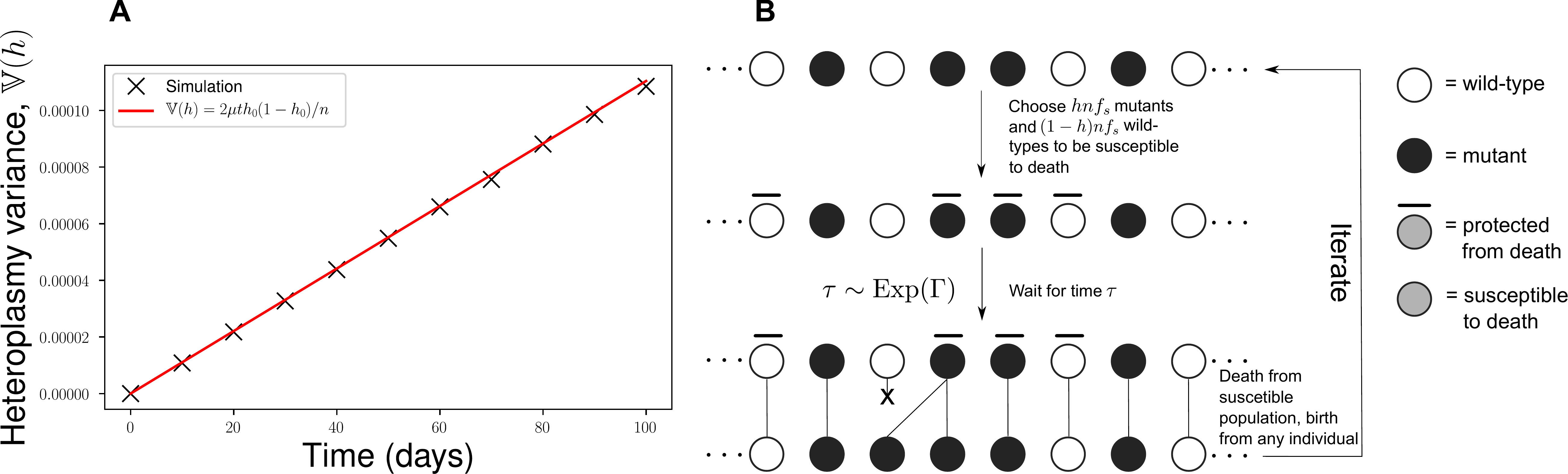}  
\end{center}
\caption{\textbf{Exploration of analogous Moran processes}. (\textbf{A})~The original biallelic Moran process satisfies Eq.~\eqref{eq:ansatz_indep_net}, where $h_0$ is the initial heteroplasmy, which is equivalent to $\mathbb{E}(h)$. (\textbf{B})~The ``protected'' Moran process. The population size is constrained to be fixed to some large constant, $n$. There exist two alleles, mutants (black circle) and wild-types (white circles). $h n f_s$ mutant and $(1-h) n f_s$ wild-type molecules are susceptible to death, the rest are protected from death (denoted by a bar). An exponential random variable is drawn as the waiting time to the next event (see \nameref{SI:mod_moran} for discussion of the form of the rate $\Gamma$). Time is incremented by the waiting time, then a death event occurs from the susceptible population and a birth event from any individual simultaneously. The same individual is allowed to be chosen for both birth and death. The process is then repeated iteratively.}
\label{Fig:moran}
\end{figure}

Suppose that, instead of the process occurring with discrete time, instead the process occurs with continuous time, where each event is a simultaneous birth and death, and is modelled as a Poisson process. Suppose that events occur at a rate $\mu$ per capita. The waiting time between successive events ($\tau$) is an exponential random variable with rate $\mu N$. Hence the expected waiting time between successive events is
\begin{equation}
\mathbb{E}(\tau)=\frac{1}{\mu n}. \label{eq:wait_moran_poiss_simple}
\end{equation}
If we take the ratio of Eq.~\eqref{eq:h_var_moran_simple} and Eq.~\eqref{eq:wait_moran_poiss_simple}, we have
\begin{equation}
\frac{\mathbb{V}(h_{t+\tau}|m_t)}{\mathbb{E}(\tau)} = \frac{2 \mu h_t (1-h_t)}{n}. \label{eq:heuristic_simple_moran}
\end{equation}
Heuristically, one could interpret Eq.~\eqref{eq:heuristic_simple_moran} as a ratio of differentials as follows. If we were to suppose that $n$ were large enough such that $\mathbb{E}(\tau)$ is very small, and $h_t$ is approximately constant ($h_0$) after a small number of events, then
\begin{equation}
\frac{\Delta \mathbb{V}(h)}{\Delta \tau} \approx \frac{\mathbb{V}(h_{t+\tau}|m_t)}{\mathbb{E}(\tau)} \implies \deriv{\mathbb{V}(h)}{t} \approx \frac{2 \mu h_0 (1-h_0)}{n}  \implies \mathbb{V}(h,t) \approx \frac{2 \mu h_0 (1-h_0)}{n} t \label{eq:simple_moran_vh}
\end{equation}
where we have replaced the inter-event time $\tau$ with physical time $t$. This result is analogous to Eq.~\eqref{eq:ansatz_indep_net} and Eq.~(12) of \cite{Johnston16}, and agrees with simulation (Figure \ref{Fig:moran}A).

Now consider the modified Moran process in Figure \ref{Fig:moran}B, which we refer to as a ``protected'' Moran process. Let $0 < f_s \leq 1$ be the fraction of individuals susceptible to death, which is a constant. $n f_s h_t$ and $n f_s (1-h_t)$ mutants and wild-types are randomly chosen to be susceptible to death, respectively, where $n$ is large. In this continuous-time model, the inter-event time is $\tau \sim \Exp(\Gamma)$ where $\Gamma$ will be defined below. Then an individual from the susceptible population is chosen for death, and any individual is allowed to be born. The birth and death events occur simultaneously in time.

Again, using $t$ as an integer counter of events, we have
\begin{eqnarray}
P(m_{t+1}=j|m_t=i)&=&
\begin{cases}
\frac{n f_s h_t}{n f_s} \frac{n (1-h_t)}{N}& \text{if } j = i - 1\\
\frac{n f_s (1-h_t)}{n f_s} \frac{n h_t}{N}& \text{if } j = i + 1\\
\frac{n f_s h_t}{n f_s} \frac{n h_t}{N} + \frac{n f_s (1-h_t)}{n f_s} \frac{n (1-h_t)}{n} & \text{if } j = i \\
0 & \text{otherwise}\\
\end{cases}\\
&=& 
\begin{cases}
h_t (1-h_t)& \text{if } j = i\pm 1\\
h_t^2 + (1-h_t)^2 & \text{if } j = i\\
0 & \text{otherwise}
\end{cases}
\end{eqnarray}
which is equivalent to the definition of a Moran process in Eq.~\eqref{eq:moran_defn}, meaning that Eq.~\eqref{eq:h_var_moran_simple} applies to the protected Moran process as well.

We consider two heuristic arguments for choosing the inter-event rate $\Gamma$, where the inter-event time $\tau \sim \Exp(\Gamma)$. Firstly, if the death rate per capita is constant ($\mu$), then the rate at which a death event occurs in the system ($\Gamma_{\text{death}}$) is proportional to the number of individuals which are susceptible to death: $\Gamma_{\text{death}} = \mu n f_s$. If we assume that the overall birth rate is matched to the overall death rate so that population size is maintained, as is the case when $\lambda = \lambda(\x)$ in the network system, then the overall birth rate ($\Gamma_{\text{birth}}$) must also be  $\Gamma_{\text{birth}} = \mu n f_s$. Hence, 
\begin{equation}
\Gamma= \Gamma_{\text{birth}} + \Gamma_{\text{death}} = 2 \mu n f_s 
\end{equation}
where $\mu$ is a proportionality constant. Since, for a Moran event to occur, both a birth and a death event must occur, time effectively runs twice as fast in a Moran model relative to a comparable chemical reaction network model. We therefore rescale time by taking $\mu \rightarrow \mu / 2$, and thus
\begin{equation}
\Gamma= \mu n f_s. \label{eq:moran_rate_ctrl_rep}
\end{equation}
As a result, $\mathbb{E}(\tau) = 1/(\mu n f_s)$ and therefore, using Eq.~\eqref{eq:h_var_moran_simple} and the reasoning in Eq.~\eqref{eq:simple_moran_vh},
\begin{equation}
\frac{\Delta \mathbb{V}(h)}{\Delta \tau} \approx \frac{\mathbb{V}(h_{t+\tau}|m_t)}{\mathbb{E}(\tau)} \implies \deriv{\mathbb{V}(h)}{t}  \approx  \frac{2 \mu f_s h_0 (1-h_0)}{n} \implies \mathbb{V}(h) \approx \frac{2 \mu f_s h_0 (1-h_0)}{n} t. \label{eq:rescaled_moran_vh}
\end{equation} 
This is analogous to when $\lambda = \lambda(\x)$ and $\mu=$ const in the network system. \textit{Hence, when $\lambda = \lambda(\x)$ and $\mu=$ const, the absence of death in the fused subpopulation means the timescale of the system (being the time to the next death event) is proportional to $f_s$}. This argument is only a heuristic, since the Moran process is defined such that birth and death events occur simultaneously and therefore do not possess separate propensities ($\Gamma_{\text{birth}}$ and $\Gamma_{\text{death}}$). 

The second case we consider is when each individual has a constant rate of birth, hence $\Gamma_{\text{birth}} \propto n$. Then the death rate is chosen such that $\Gamma_{\text{birth}}=\Gamma_{\text{death}}$. In this case $\Gamma = \lambda n$, where $\lambda$ is a proportionality constant. The same argument from Eq.~\eqref{eq:wait_moran_poiss_simple} to Eq.~\eqref{eq:simple_moran_vh} may be applied, with an appropriate rescaling of time, and we arrive at Eq.~\eqref{eq:ansatz_indep_net}. This is analogous to when $\mu = \mu(\x)$ and $\lambda = $ const in the network system. \textit{Hence, when $\mu = \mu(\x)$ and $\lambda = $ const, the presence of a constant birth rate in the entire population means the timescale of the system (being the time to the next birth event) is independent of $f_s$}.

\newpage

\bgroup
\def\arraystretch{1.5}
\begin{table}[h!]
\caption{\textbf{Key predictions from our mathematical models}. } \label{tab:key_findings}
\begin{tabularx}{\textwidth}{|X|}
\hline
The following results hold for our neutral genetic model of a post-mitotic cell, with a simple model of mitochondrial network dynamics:
\begin{enumerate}
\item The rate of increase of heteroplasmy variance is proportional to the fraction of unfused mitochondria, but independent of the absolute magnitude of fission-fusion rates, due to a rescaling of time by the mitochondrial network (Eq.\eqref{eq:ansatz}, Eq.\eqref{eq:SDE_network_sys}, Figure \ref{Fig:model_expl}E-H).
\item The rate of accumulation of \textit{de novo} mutations increases as the fraction of unfused mitochondria increases, due to a rescaling of time by the mitochondrial network (Figure \ref{Fig:inf_sites}B-D)
\item When fusion is selective, intermediate fusion-fission ratios are optimal for reducing mean heteroplasmy (Figure \ref{Fig:qc}A)
\item When mitophagy is selective, complete fission is optimal for reducing mean heteroplasmy (Figure \ref{Fig:qc}B)
\end{enumerate}\\
\hline
\end{tabularx}
\end{table}
\egroup

\bgroup
\def\arraystretch{1.5}
\begin{table}[h!]
\caption{\textbf{Nominal parametrization for network system}. Dimensions of parameters are derived using individual terms in Eqs.~\eqref{eq:phi_ws}--\eqref{eq:phi_mf}; copy numbers of particular species are pure numbers and therefore dimensionless. Since we have chosen a system size of $\Omega=1$ throughout, we set the dimension of volume to 1. See \nameref{SI:nom_param} for further details of parameter justification.}
\label{table:params}
\begin{tabularx}{\textwidth}{lXlXX}
\textbf{Parameter} & \textbf{Description} & \textbf{Value} & \textbf{Dimensions}& \textbf{Remarks} \\
\hline
$\beta$  &  Fission rate  &  33.12  &  day$^{-1}$  & For cortical neurons, see \citep{Cagalinec13}   \\
$\gamma$ & Fusion rate    &  $3.79\times10^{-2}$ & day$^{-1}$ & Approximately balances fission \\
$\mu$  & Mitophagy rate  & 0.023   &  day$^{-1}$  & From \citep{Tam15} \\
$\delta$  & Mutant feedback sensitivity  & 1   & dimensionless & Potentially appropriate for e.g.\ non-coding mtDNA mutations \\
$b$ & Feedback control strength & $1.24\times10^{-5}$ & day$^{-1}$ & Chosen as the weakest control strength which has a non-trivial steady state and total copy number of 1000 \\
$\kappa$ & Steady state copy number parameter & 11.7 & dimensionless & See remark for $b$
\end{tabularx}
\end{table}
\egroup

\newpage

\begin{figure}[h!]
\begin{center}
\includegraphics[width=0.7\columnwidth]{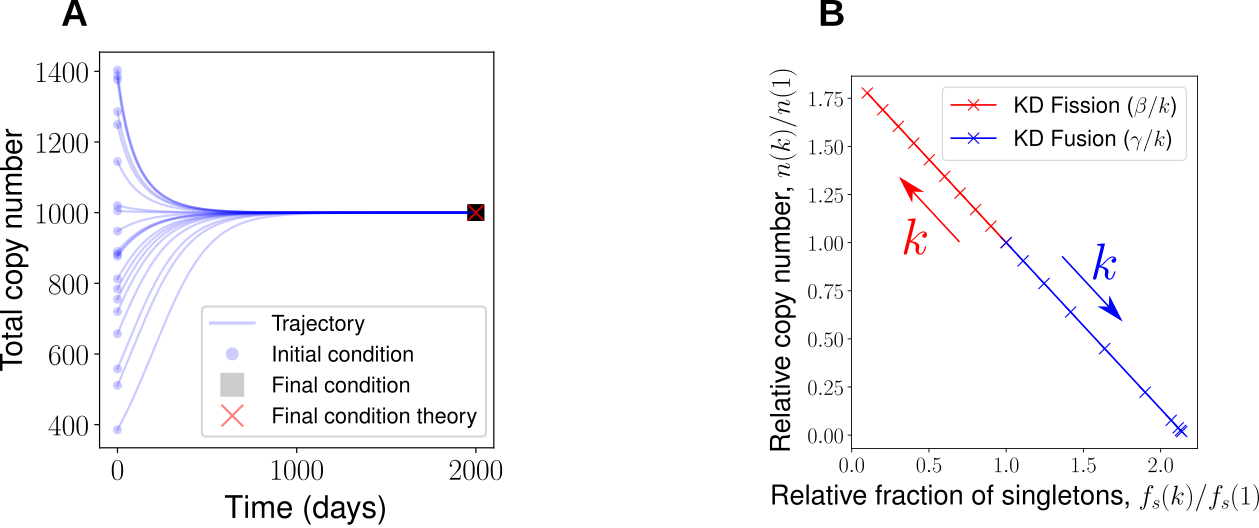}  
\end{center}
\caption{\textbf{Deterministic treatment of network system}. (\textbf{A})~Deterministic dynamics of total copy number under linear feedback control, which is controlled to a particular steady state value (see Figure~\ref{Fig:model_expl}A\&{}B).  (\textbf{B})~Defining a knock-down (KD) factor ($k^{-1}=0.1,0.2\dots,1.0$), the fission rate was rescaled to $\beta \rightarrow \beta/k$ (red) and the fusion rate to $\gamma \rightarrow \gamma/k$ (blue), causing a linear increase and decrease in total copy number respectively under a deterministic treatment (see Figure~\ref{Fig:model_expl}A\&{}B). }
\label{Fig:det_extras}
\end{figure}

\newpage

\begin{figure}[h!]
\begin{center}
\includegraphics[width=0.7\columnwidth]{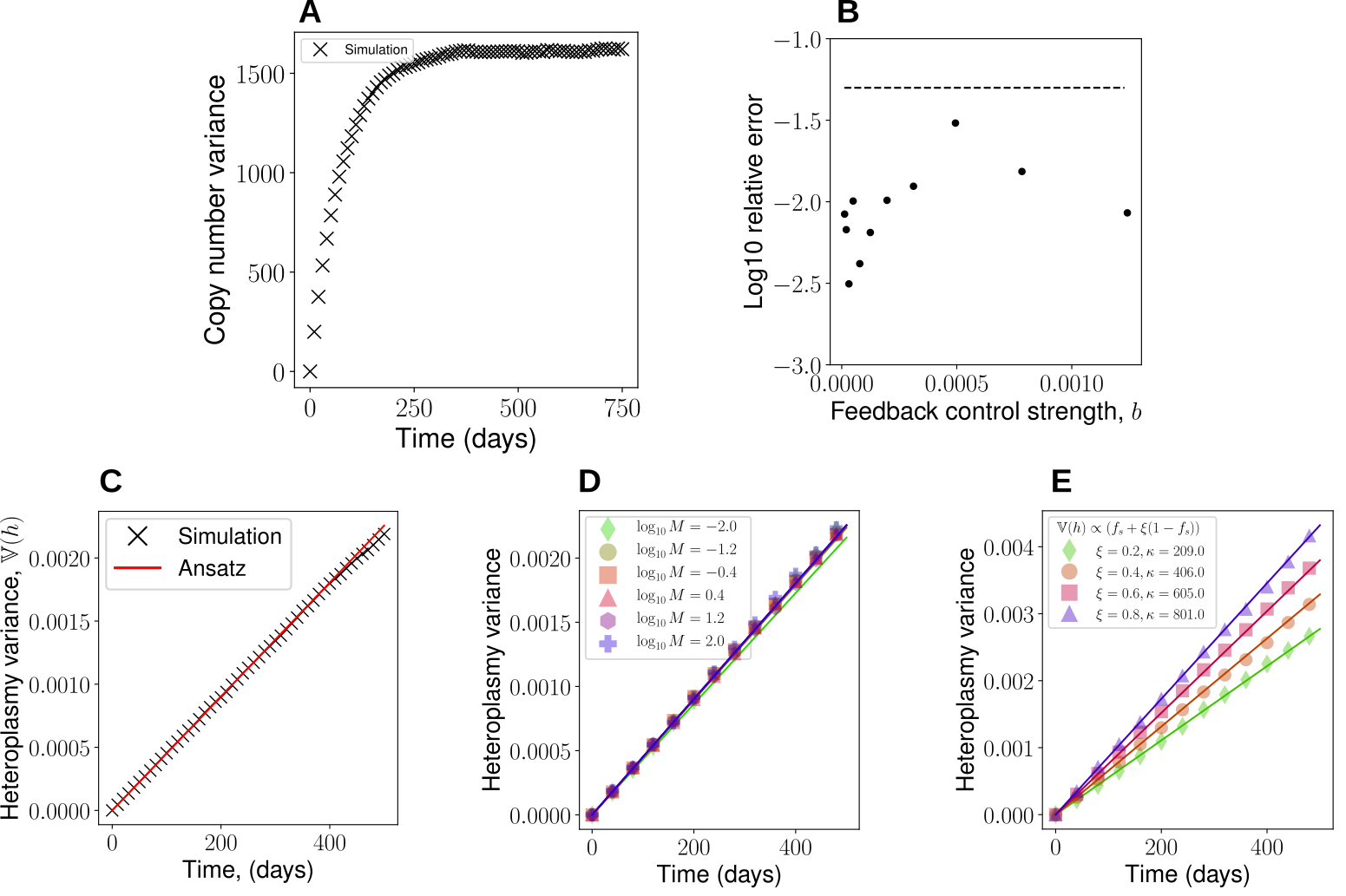}  
\end{center}
\caption{\textbf{Stochastic treatment of network system}. (\textbf{A})~Copy number variance for stochastic simulations initially increases, since all stochastic simulations begin with the same initial condition, but then plateaus since the steady state line is globally attracting (see Figure~\ref{Fig:model_expl}C\&{}D). (\textbf{B})~Error in Eq.~\eqref{eq:ansatz} in a sweep over the feedback control strength, $b$. Dotted line denotes a 5\% error according to Eq.~\eqref{eq:error_metric}. (\textbf{C})~$\mathbb{V}(h)$ profile for the parametrization with the largest error in (\textbf{B}). (\textbf{D})~Sweeps of the network rate magnitude (see Figure~\ref{Fig:model_expl}H). Heteroplasmy variance is approximately independent of absolute network rates over a broad range of network magnitudes. (\textbf{E})~Allowing fused species to be degraded with relative rate $\xi$ (Eq.~\eqref{eq:fused_deg_xi}), stochastic simulations for heteroplasmy variance (markers) and Eq.~\eqref{eq:ansatz_xi} (lines) are shown. Fused species degradation induces a linear correction to the heteroplasmy variance formula. }
\label{Fig:stoch_extras}
\end{figure}

\newpage

\begin{figure}[h!]
\begin{center}
\includegraphics[width=0.8\columnwidth]{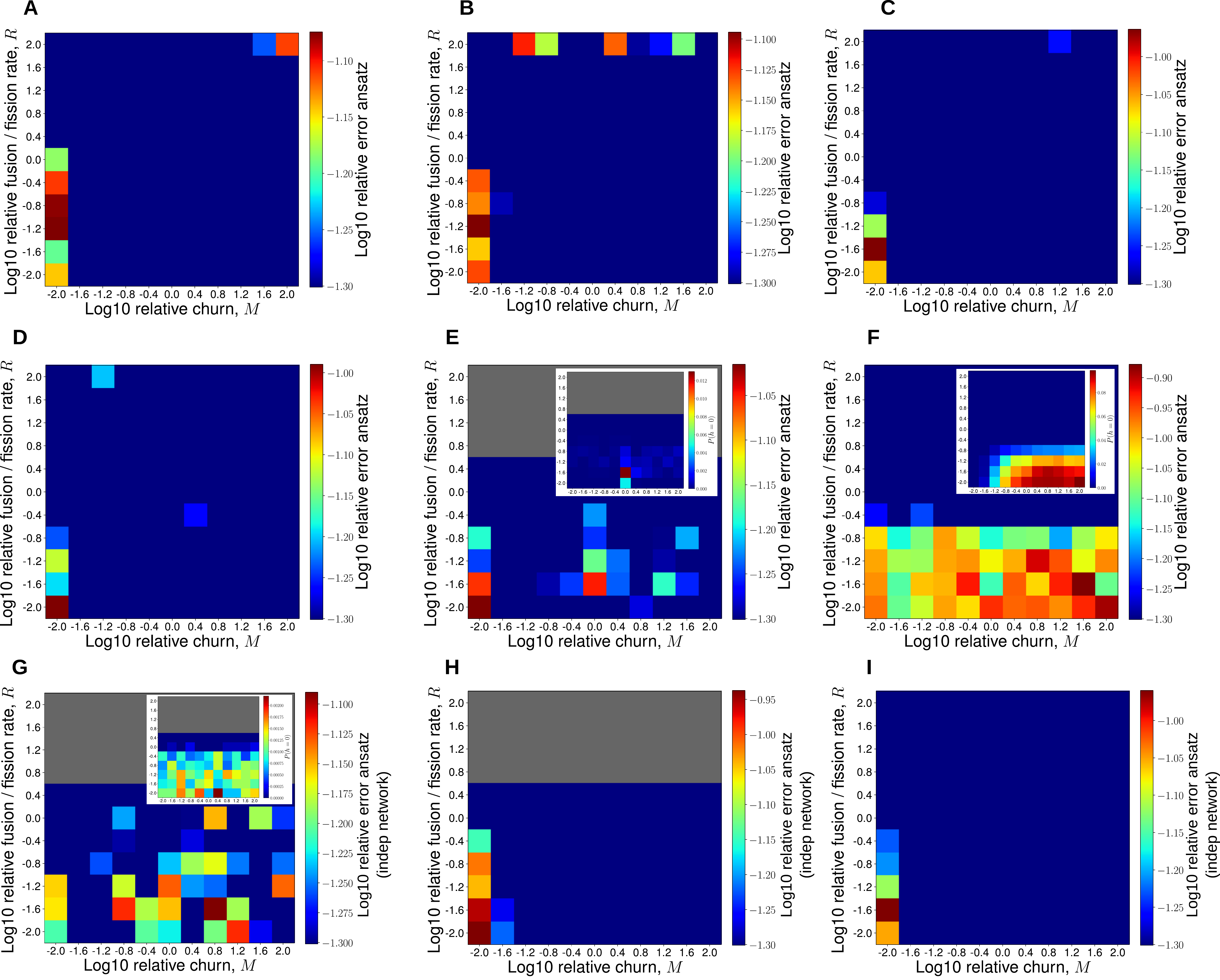}  
\end{center}
\end{figure}
\begin{center}
\captionof{figure}{ \textbf{Parameter sweeps of network fission-fusion rates for replication-based and degradation-based control modes show robustness of two heteroplasmy variance formulae respectively}. (\textbf{A-I})~Error in Eq.~\eqref{eq:ansatz} (ansatz) and Eq.~\eqref{eq:ansatz_indep_net} (ansatz indep network)  for sweeps of the fusion and fission rates for the corresponding feedback control functions in Table \ref{tab:param_sweeps}. Equations are accurate to at least 5\% (blue regions) across large regions of parameter space, for many control laws. Fusion and fission rates are redefined as $\gamma \rightarrow \gamma_0 M R$ and $\beta \rightarrow \beta_0 M$ where $M$ and $R$ denote the magnitude and ratio of the network rates, and $\gamma_0, \beta_0$ denote the nominal parametrizations of the fusion and fission rates respectively. Summary statistics for $10^4$ realizations, with initial condition $h=0.3$ and evaluated at $t=500$~days. Errors in $\mathbb{V}(h)$ (see Eq.~\eqref{eq:error_metric}) smaller than $5\%$ are truncated and are shown as blue. Parametrizations where a deterministic steady state could not be found for an initial condition of $h=0.3$ are shown in grey. Inset figures, where present, display the probability of fixation at $h=0$. Where insets are not present, the probability of fixation is negligible. (\textbf{A-F}) When $\lambda = \lambda(x)$ and $P(h=0)$ is low, Eq.~\eqref{eq:ansatz} performs well in the high-churn limit. (\textbf{G-I}) When $\mu = \mu(x)$ and $P(h=0)$ is low, Eq.~\eqref{eq:ansatz_indep_net} performs well in the high-churn limit. }
\label{Fig:sweeps}
\end{center}

\newpage

\bgroup
\def\arraystretch{1.5}
\begin{table}[h!]
\caption{\textbf{Nominal parametrizations for the alternative feedback control functions explored}. Nominal parametrizations for the feedback controls in Fig.~\ref{Fig:sweeps}. In all cases, the nominal fission and fusion rates were $\beta=33.12$, $\gamma = 0.038$ respectively.} \label{tab:param_sweeps}
\begin{tabularx}{\textwidth}{X|X|X|X}
\textbf{Interpretation} & \textbf{Replication rate} & \textbf{Degradation rate} & \textbf{Note} \\
\hline
Linear feedback (see Figure~\ref{Fig:model_expl}, \ref{Fig:det_extras}, \ref{Fig:stoch_extras}, \ref{Fig:sweeps}A) & $\mu + b (\kappa - w_T - \delta m_T)$; see Table \ref{table:params} & $\mu$; see Table \ref{table:params} & Control E in (Johnston and Jones, 2016) and GitHub repository \\ \hline
Relaxed replication (see Figure \ref{Fig:sweeps}B) & $\alpha \mu (w_{opt} - w_T - \delta m_T)/(w_T+m_T)$; $\alpha=1$, $w_{opt} = 1000$, $\delta = 1$ & $\mu$; $\mu = 0.023$ & Control A in (Johnston and Jones, 2016) and GitHub repository \\ \hline
Differential control for target population (see Figure \ref{Fig:sweeps}C) & $\alpha (w_{opt}-w_T)$ & $\mu$; $\mu = 0.023$ & Control B in (Johnston and Jones, 2016) and GitHub repository  \\ \hline
Ratiometric control for target population  (see Figure \ref{Fig:sweeps}D) & $\alpha (w_{opt}/w_T - 1)$; $\alpha=1$, $w_{opt}=1000$ & $\mu$; $\mu = 0.023$ & Control C in (Johnston and Jones, 2016) and GitHub repository \\ \hline
Production independent of wild-type (see Figure \ref{Fig:sweeps}E) & $\alpha /w_T$; $\alpha=5$ & $\mu$; $\mu = 0.023$ & Control F in (Johnston and Jones, 2016) and GitHub repository \\ \hline
General linear feedback (see Figure \ref{Fig:sweeps}F) & $\mu + b (\kappa - \delta_1 w_s - \delta_2 w_f - \delta_3 m_s - \delta_4 m_f)$; see Table \ref{table:params}, $\delta_1 = 0.8$, $\delta_2 = 1.0$, $\delta_3 = 0.2$, $\delta_4 = 0.3$ & $\mu$; see Table \ref{table:params} & Control X in GitHub repository \\ \hline
Ratiometric control through degradation (see Figure \ref{Fig:sweeps}G) & $\lambda$; $\lambda = 0.023$ & $\mu w_T / w_{opt}$; $\mu = 0.023$; $w_{opt}=200$ & Control G in (Johnston and Jones, 2016) and GitHub repository \\ \hline
Linear feedback in degradation (see Figure \ref{Fig:sweeps}H) & $\lambda$; $\lambda = 0.023$ & $\mu + b (w_T + \delta m_T - \kappa)$; $\mu = 0.023$, $b=10^{-4}$, $\kappa = 1000$, $\delta = 1$& Control Y in GitHub repository \\ \hline
Differential control for target population in degradation (see Figure \ref{Fig:sweeps}I) & $\lambda$; $\lambda = 0.023$ & $\alpha (w_T - w_{opt})$; $\alpha = 1$, $w_{opt}=1000$& Control Z in GitHub repository \\ 
\end{tabularx}
\end{table}
\egroup

\newpage

\begin{figure}[h!]
\begin{center}
\includegraphics[width=0.5\columnwidth]{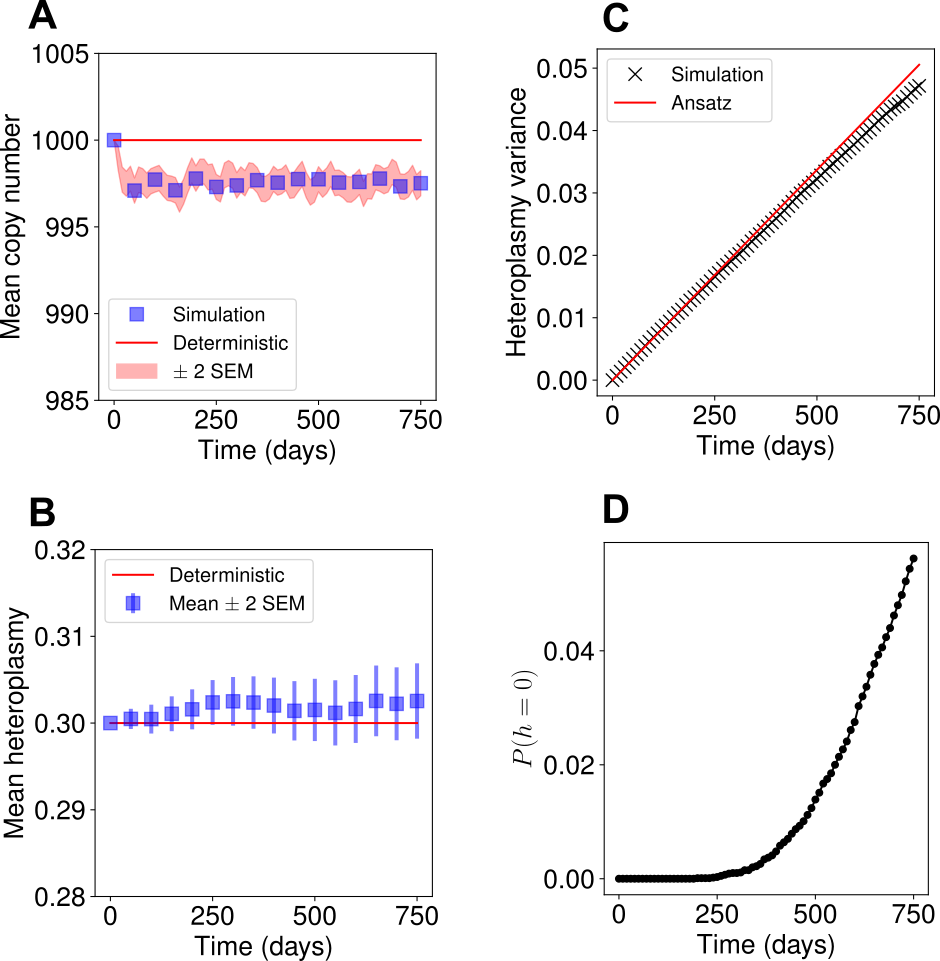}  
\end{center}
\caption{\textbf{Ansatz predicts heteroplasmy variance for linear feedback control in a fast mtDNA turnover regime when fixation probability is low}. Stochastic simulations of the linear feedback control network system with an mtDNA half-life of 2~days \citep{Poovathingal12}, corresponding to $\mu = \ln(2)/2$. (\textbf{A}) The mean copy number, and (\textbf{B}) the mean heteroplasmy show qualitatively similar dynamics to the nominal parametrisation presented (see Fig.~\ref{Fig:model_expl}C\&{}D ). (\textbf{C}) Eq.\eqref{eq:ansatz} predicts $\mathbb{V}(h)$ accurately up to approximately 250~days, where the Eq.\eqref{eq:ansatz} begins to overestimate the variance. (\textbf{D}) The over-estimation of heteroplasmy variance coincides with an increase in the probability of fixation at $h=0$. Parameters apart from $\mu$ were chosen according to the protocol outlined in \nameref{SI:nom_param}, with $10^4$ repeats. $\kappa = 101.6$, $b=2.07\times10^{-4}$, for all other parameters see Table~\ref{table:params}.}
\label{Fig:lfc_low_half_life}
\end{figure}

\newpage

\begin{figure}[h]
\begin{center}
\includegraphics[width=0.5\columnwidth]{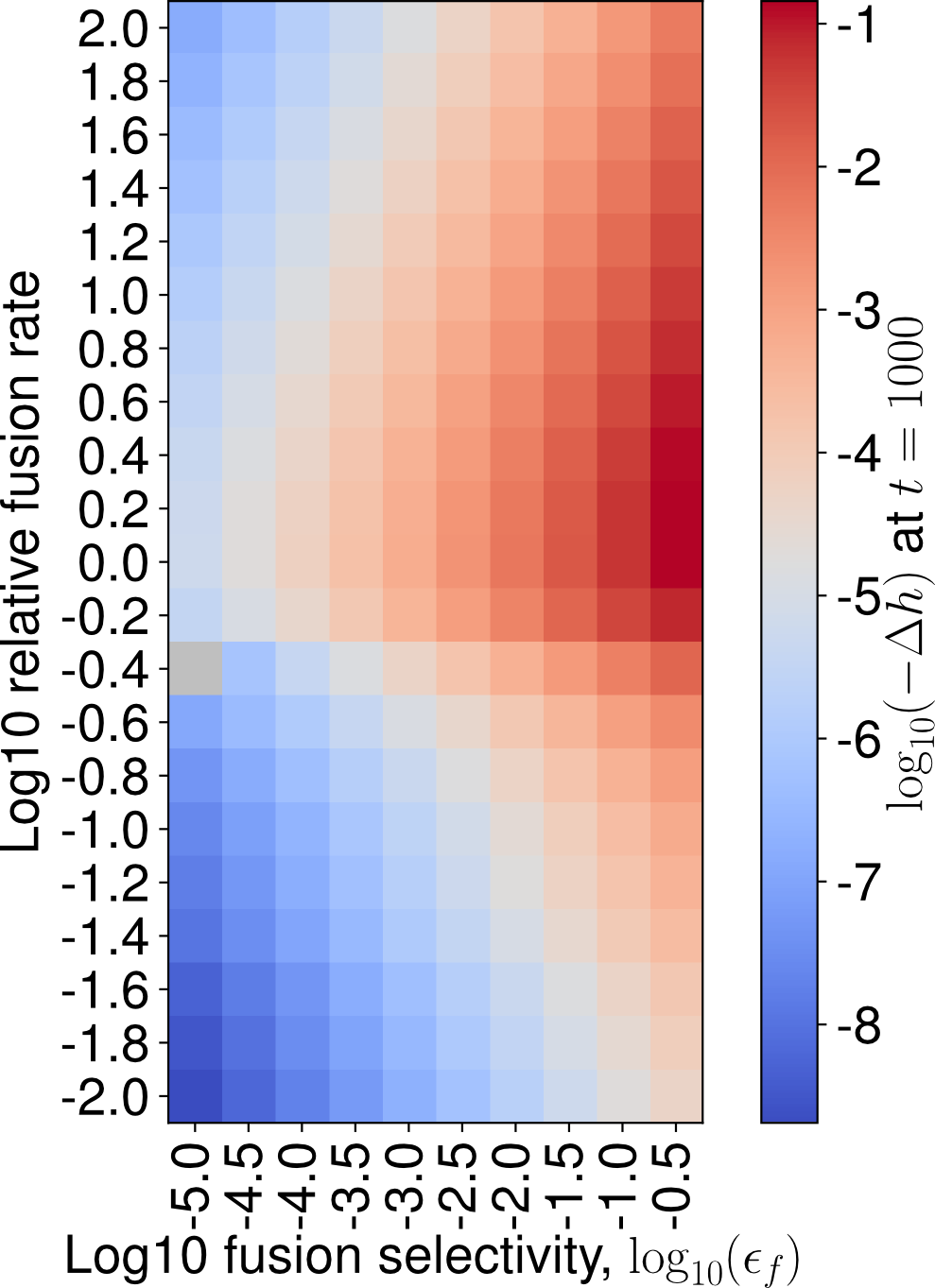}  
\end{center}
\caption{\textbf{A deterministic parameter sweep of fusion selectivity and the relative fusion rate for mitochondrial quality control}. An ODE treatment allows smaller heteroplasmy changes to be probed without the need to resort to an infeasible number of stochastic simulations. Displaying the relative change in heteroplasmy ($\Delta h$) after $t=1000$ days. We observe that a reduction in heteroplasmy is achieved at intermediate fusion rates at all non-zero fusion selectivities investigated. Grey denotes a change which is smaller than floating point precision.}
\label{Fig:QC_det_sweep}
\end{figure}

\bgroup
\def\arraystretch{1.5}
\begin{table}[h!]
\caption{\textbf{Comparison of previous models with our model of mitochondrial genetic/network dynamics}. \ud{Each of the key differences with our model is enumerated, and has a corresponding comment;} see \nameref{sec:discussion}. } \label{tab:comparison_models}
\begin{tabularx}{\textwidth}{X|X|X|X}
\textbf{Model} & \textbf{Key assumptions} & \textbf{Key differences} & \textbf{Comments}\\
\hline
\citep{Tam13,Tam15} & \begin{itemize}
\item Cell consists of 16 sub-compartments 
\item Fission/fusion induces migration between subcompartments
\end{itemize}  & \begin{enumerate}
\item Slower fission-fusion dynamics result in larger rate of increase in $\mathbb{V}(h)$
\item Fast fission-fusion rates cause $\mathbb{E}(h)$ to increase
\end{enumerate} & \begin{enumerate}
\item Limiting to regimes where fission-fusion is fast, likely results in loss of rate magnitude sensitivity
\item Could be explained by spatial effects which we neglect here
\end{enumerate} \\
\hline
\citep{Mouli09} & \begin{itemize}
\item Fission follows fusion
\item Mitochondria consist of multiple units
\item Sigmoidal relationship between number of functional units per mitochondrion and activity
\end{itemize} & \begin{enumerate}
\item When fusion is selective, higher fusion rates are optimal
\item When mitophagy is selective, intermediate fusion rates are optimal
\end{enumerate} & \begin{enumerate}
\item The optimal fusion rate is an order of magnitude lower than the highest fusion rate considered
\item Non-linearity between function, intra-mitochondrial heteroplasmy, and network state, may result in intermediate fusion optimality
\end{enumerate} \\
\hline
\end{tabularx}
\end{table}
\egroup

\end{document}